\documentclass[12pt]{iopart}
\usepackage{times}
\usepackage{epsfig}
\usepackage{marvosym}

\def\lam{$\lambda_{\rm int}$}

\def\NIM{Nucl. Instr. and Meth.~}
\def\pl{Phys. Lett.~}
\def\ieee {{IEEE Trans. Nucl. Sci.~}}

\def\etal{{\it et al.}}
\def\eg{{\it e.g.,~}}
\def\ie{{\it i.e.,~}}

\def\etc{{\it etc.~}}
\def\vs{{\it vs.~}}

\begin{document}

\title{Misconceptions about Calorimetry}
\author{Michele Livan$^1$ and Richard Wigmans$^2$}
\address{%
$^{1}$ \quad Dipartimento di Fisica, Universit\`a di Pavia and INFN Sezione di Pavia, Via Bassi 6, Pavia 27100 Italy; michele.livan@unipv.it\\
$^{2}$ \quad Department of Physics, Texas Tech University, Lubbock, TX 79409-1051, U.S.A.; wigmans@ttu.edu}

\begin{abstract}
{In the past 50 years, calorimeters have become the most important detectors in many particle physics experiments,
especially experiments in colliding-beam accelerators at the energy frontier. In this paper, we describe and discuss
a number of common misconceptions about these detectors, as well as the consequences of these misconceptions.
We hope that it may serve as a useful source of information for young colleagues who want to familiarize themselves 
with these tricky instruments.}
\end{abstract}

\maketitle

\section{Introduction}
\vskip 2mm
In the past fifty years, calorimeters have become a very important component of the toolbox of particle physicists.
Especially in experiments at large storage rings, where beams of high-energy particles are brought into collision with each
other, calorimeter systems are the heart and soul of the detector system. Some of the  reasons that make calorimeters
extremely suitable for such experiments are
\begin{enumerate}
\item The fact that they are sensitive to both the charged and neutral particles produced in the interactions.
\item The fact that their performance typically improves with increasing energies.
\vskip 1mm
These two characteristics distinguish them
from other detector components, which are typically only sensitive to charged particles and become less performant at increasing
energy.
\vskip 1mm
\item The fact that they can provide the information needed for deciding whether a certain event is worth retaining for 
further (offline) inspection {\sl extremely fast}, \ie almost instantaneously.
\vskip 1mm
This is the most decisive reason for the important role
of calorimeters in modern experiments, where the interesting events sometimes (\ie at CERN's Large Hadron Collider)
represent a very tiny fraction of the total. The capability of calorimeter systems to provide information on the {\sl energy flow} in the
events (missing energy, transverse energy, jet production, \etc) is an extremely valuable feature in this context.
\end{enumerate}
Despite the crucial role of calorimeters in many experiments, there are unfortunately still many misconceptions about
these instruments. These misconceptions derive from the, in many ways, counter-intuitive performance characteristics.
They have led and continue to lead to fundamental mistakes in designing detector systems, and interpreting results of
experiments.

In this paper, we review a few common misconceptions, and describe some of the practical consequences. We mainly concentrate on 
{\sl sampling} calorimeters, since that is where most of the problems occur. With few exceptions, most calorimeter systems at 
large $4\pi$ experiments at colliders use indeed different materials for the absorption of the particles and for generating the signals
resulting from this absorption process. 
Section 2 contains a brief introduction to calorimetry as a particle detection technique. In Section 3, 
the misconceptions and their consequences are described. In Section 4, we show how beam tests of prototype 
calorimeter modules are often at the origin of the problems. Conclusions and our outlook for the future are given in Section 5.

\section{Calorimetry as a particle detection technique}

In nuclear and particle physics, the term calorimetry refers to the detection of particles, and
measurement of their properties, through total absorption in an instrument called
a {\sl calorimeter}.
Calorimeters exist in a wide variety, but they all have the common feature that 
the measurement process through which the particle properties are determined is {\sl destructive}.
Unlike, for example, wire chambers that measure a particle's properties by tracking it in
a magnetic field, the particles are no longer available for inspection by other devices once
the calorimeter is done with them. The only exception to this rule concerns muons. The fact
that these particles may penetrate the substantial amount of matter represented by a 
calorimeter without losing much of their energy is actually an important ingredient for their identification as muons. Other particles (neutrinos and particles hypothesized in the context of Supersymmetry) do not leave any trace in a calorimeter, or in any other detector component. Yet, calorimeters are also crucial
tools for recognizing the presence of these particles, and measuring their properties.

In the absorption process, which is usually called {\sl shower development}, almost all the particle's energy is eventually converted into heat,
hence the term calorimetry. However, the units of the energy involved in this process are typically 
very different from the thermodynamic ones. The most energetic particles in modern accelerator
experiments are measured in units of TeV (1 TeV = $10^{12}$ eV = 1000 GeV), whereas 1 calorie (4.18 Joule)
is equivalent to about $10^7$ TeV.
The rise in temperature of the particle detector is thus, for all practical purposes,
negligible, and therefore other ways to measure the deposited energy are employed. These methods are typically
based on the measurable effects of atomic or molecular excitation (ionization charge, scintillation light), or on
collective effects such as the production of \v{C}erenkov light or sound in the absorbing medium. 

\subsection{Functions and properties of calorimeters}

Calorimeters measure the energy released in the absorption of (sub)nuclear particles that enter them. 
They generate signals that make it possible to quantify that energy. Typically, these signals
provide also other information about the particles, and about the event in which they were produced.
The signals from a properly instrumented absorber may be used to measure the entire four-vector of the particles. 

By analyzing the energy deposit pattern, the direction of the particle can be measured.
The mass of the showering particle can be determined in a variety of ways, \eg from the time structure of the signals, the energy deposit profile, or a comparison of the measured energy and momentum of the particle. Calorimeters are also used to identify muons and neutrinos. High-energy muons usually deposit only a small fraction of their energy in the calorimeter and produce signals in downstream detectors. Neutrinos typically do not interact at all in the calorimeter. If an energetic neutrino is produced in a colliding-beam experiment, this phenomenon will lead to an imbalance between the energies deposited in any two hemispheres into which a $4\pi$ detector can be split.
Such an imbalance is usually referred to as {\sl missing transverse energy}.

The latter is an example of the {\sl energy flow information} a calorimeter system can provide.
Other examples of such information concern the {\sl total transverse energy} and the production of {\sl hadronic jets}
in the measured events. Since this information is often directly related to the physics goals of the experiment, and since it can be obtained extremely fast, calorimeters usually play a crucial role in the trigger scheme, through which interesting events are selected and retained for further inspection off-line. 

The calorimeter's properties should be commensurate with the role it has to play in the experiment.
Relevant properties in this context are the energy resolution, the size (which determines the effects of shower leakage), the signal speed and the hermeticity.

\subsection{Calorimeter types}

One frequently distinguishes between {\sl homogeneous} and {\sl sampling} calorimeters.
In a homogeneous calorimeter, the entire detector volume is sensitive to the particles and may 
contribute to the generated signals.
In a sampling calorimeter, the functions of
particle absorption and signal generation are exercised by {\sl different} materials, called 
the {\sl passive} and {\sl active medium}, respectively. The passive medium is usually a high-density
material, such as iron, copper, lead or uranium. The active medium generates the light or charge
that forms the basis for the signals from such a calorimeter.

In some non-accelerator experiments, the calorimeter is also the {\sl source} that generates the particles
to be detected. As examples, we mention large water \v{C}erenkov counters built to detect astrophysical neutrinos and the high-purity $^{76}$Ge crystals or the $^{136}$Xe liquid used to study $\beta \beta$ decay.

\subsubsection{\sl Electromagnetic calorimeters}

Electromagnetic calorimeters are specifically intended for the detection of energetic electrons and 
$\gamma$s, but produce usually also signals when traversed by other types of particles. They are used over a very wide energy range, from the semiconductor crystals that measure $X$-rays down to a few keV to shower counters such as AGILE, PAMELA and FERMI, which orbit the Earth on satellites in search for electrons, positrons and $\gamma$s with energies $>10$ TeV.
These calorimeters don't need to be very deep, especially when high-$Z$ absorber material is used. For example, when 100 GeV electrons enter a block of lead, $\sim 90\%$ of their energy is deposited in only 4 kg of material.
%
By far the best energy resolutions have been obtained with large semiconductor crystals, and in particular high-purity germanium. These are the detectors of choice in nuclear $\gamma$ ray spectroscopy, and routinely obtain resolutions ($\sigma/E$) of 0.1\% in the 1 MeV energy range. The next best class of detectors are scintillating crystals, which are often the detectors of choice in experiments involving $\gamma$ rays in the energy range from
1 - 20 GeV, which they measure with energy resolutions of the order of 1\%. Excellent performance in this energy range has also been reported for liquid krypton and xenon detectors, which are bright (UV) scintillators. Other homogeneous detectors of em showers are based on \v{C}erenkov light, in particular 
lead glass. Very large water \v{C}erenkov calorimeters (\eg SuperKamiokande) should also be mentioned in this category.

Sampling calorimeters, which are typically much cheaper, become competitive at higher energies.
In properly designed instruments of this type, the energy resolution is determined by {\sl sampling fluctuations}. 
These represent fluctuations in the number of different shower particles that contribute to the  calorimeter signals, convoluted with fluctuations in the amount of energy deposited by individual shower particles in the active calorimeter layers. They depend both on the {\sl sampling fraction}, which is determined by  the ratio of active and passive material, and on the {\sl sampling frequency}, determined by the number of different sampling elements in the region where the showers develop. 
Sampling fluctuations are stochastic and their contribution to the energy resolution is described by \cite{Liv95}
\vskip 1mm
\begin{equation}
{\sigma \over E}~= {a \over \sqrt{E}}~~~~~~{\rm with} ~~~~~a~=~0.027 \sqrt{d/f_{\rm samp}}
\label{eq:reso}
\end{equation}
in which $d$ represents the thickness of individual active sampling layers (in mm), and $f_{\rm samp}$ the sampling fraction for minimum ionizing particles ({\sl mip}s).
This expression describes data obtained with a large variety of different (non-gaseous) sampling calorimeters reasonably well.

\subsubsection{\sl Hadron calorimeters}

The energy range covered by hadron calorimeters is in principle even larger than that for em ones. Calorimetric techniques are
used to detect thermal neutrons, which have kinetic energies of a small fraction of 1 eV, to the highest-energy particles
observed in nature, which reach the Earth from outer space as cosmic rays carrying up to $10^{20}$ eV or more. In accelerator-based particle physics
experiments, hadron calorimeters are typically used to detect protons, pions, kaons and fragmenting quarks and gluons (commonly referred to as {\sl jets}) with energies in the GeV - TeV range. In this paper, we mainly discuss the latter instruments.

The development of hadronic cascades in dense matter differs in essential ways from that of electromagnetic ones, with important consequences for calorimetry.
Hadron showers consist of two distinctly different components:
\begin{enumerate}
\item An {\sl electromagnetic} component; $\pi^0$s and $\eta$s generated in the absorption process
decay into $\gamma$s which develop em showers.
\item A {\sl non-electromagnetic} component, which combines essentially everything else that
takes place in the absorption process. 
\end{enumerate}
For the purpose of calorimetry, the main difference between these components is that
some fraction of the energy contained in the non-em component does {\sl not} contribute to the signals. This {\em invisible energy}\index{Invisible energy}, which mainly consists of the binding energy of nucleons released in the numerous 
nuclear reactions, may represent up to 40\% of the total non-em energy, with large event-to-event fluctuations.

The appropriate length scale of hadronic showers is the nuclear interaction length ($\lambda_{\rm int}$)\index{Nuclear interaction length}, 
which is typically much larger (up to 30 times for high-$Z$ materials) than the radiation length ($X_0$), which governs the development of em showers.
Many experiments make use of this fact to distinguish between electrons and hadrons on the basis of the energy deposit profile
in their calorimeter system. Since the ratio $\lambda_{\rm int}/X_0$ is proportional to $Z$, particle identification on this basis works best for high-$Z$ absorber materials. Lead and depleted uranium are therefore popular choices for the absorber material in preshower detectors and the first section of a longitudinally segmented calorimeter,
which is therefore commonly referred to as the {\sl electromagnetic section}.

Just as for the detection of  em showers, high-resolution hadron calorimetry requires an average longitudinal containment better than 99\%. 
In iron and materials with similar $Z$, which are most frequently used for hadron calorimeters, 99\% longitudinal containment requires a thickness ranging from $5 \lambda_ {\rm int}$ at 20 GeV to $8 \lambda_{\rm int}$ at 150 GeV.  Hadronic energy resolutions of 1\%
require not only longitudinal shower containment at the 99\% level, but also lateral containment of 98\% or better.

Energetic $\pi^0$s may be produced throughout the absorber volume, and not exclusively in the em calorimeter section. They lead to local regions of highly concentrated energy deposit. Therefore,
there is no such thing as a ``typical hadronic shower profile". This feature affects not only the shower containment requirements, but also the
calibration of longitudinally segmented calorimeters in which one tries to improve the quality of calorimetric energy
measurements of jets with an upstream tracker, which can measure the momenta of the charged jet constituents with great precision. 
This method has become known as {\sl Particle Flow Analysis} (PFA).

\subsubsection{{\sl Compensation}}

The properties of the em shower component have also important consequences for the {\em energy resolution}, the signal {\em linearity} and the {\em response function}.
The average fraction of the total shower energy contained in the em component has been measured to increase
with energy following a power law:
\begin{equation}
\langle f_{\rm em} \rangle ~=~1 - \bigl[E/E_0\bigr]^{k-1}
\label{femE}
\end{equation}
where $E_0$ is a material dependent constant related to the average multiplicity in hadronic interactions (varying from 0.7 GeV to 1.3 GeV for $\pi$-induced reactions on Cu and Pb, respectively), and $k \sim 0.82$. For proton-induced reactions, $\langle f_{\rm em} \rangle$ is typically considerably smaller, as a result
of baryon number conservation in the shower development.

Let us define the calorimeter {\em response} as the conversion efficiency from deposited energy to generated signal, and normalize it to electrons. The responses of a given calorimeter to the em and non-em hadronic shower components, $e$ and $h$, are usually not the same, as a result of invisible energy and a variety of other effects. Such calorimeters are called {\em non-compensating} ($e/h \ne 1$)\index{Compensation}.
Since their response to hadrons, $ \langle f_{\rm em} \rangle + \bigl[1 -  \langle f_{\rm em} \rangle\bigr] h/e$, is energy dependent (\ref{femE}), they are intrinsically non-linear.

Event-to-event fluctuations in $f_{\rm em}$ are large and non-Poissonian. If $e/h \ne 1$, these fluctuations 
tend to dominate the hadronic energy resolution and their asymmetric distribution characteristics are reflected in the 
response function.
The effects  of non-compensation on resolution, linearity and line shape, as well as the associated calibration problems \cite{Olga} are absent in compensating calorimeters ($e/h = 1.0$). Compensation can be achieved in sampling calorimeters with high-$Z$ absorber material and hydrogenous active material. It requires a very specific sampling fraction, so that the response to shower neutrons is boosted by the precise factor needed to equalize $e$ and $h$. For example, in Pb/scintillating-plastic structures, this sampling fraction is $\sim 2\%$ for showers \cite{Bern87,Aco91c,Suz99}. This small
sampling fraction sets a lower limit on the contribution of sampling fluctuations, while the need to efficiently detect MeV-type neutrons requires signal integration over a relatively large volume during at least 30 ns. Yet, calorimeters of this type currently hold the world record for hadronic energy resolution ($\sigma/E \sim 30\%/\sqrt{E}$ \cite{Aco91c}).

Excellent hadronic performance has also been achieved with calorimeters that use the {\sl dual-readout method} (DREAM) \cite{Akc05}. Such calorimeters produce two signals that provide complementary information about the shower development. Since \v{C}erenkov light is almost exclusively produced in the em shower component, a comparison of the \v{C}erenkov signal with a signal to which {\sl all} charged shower particles contribute, the value of $f_{\rm em}$ can be measured for each individual event. This makes it possible to eliminate the detrimental effects of fluctuations in this variable and achieve similar performance as intrinsically compensating calorimeters, without the mentioned disadvantages.

\section{Common misconceptions and their consequences}
\vskip 2mm
\subsection{Shower particles contributing to the calorimeter signals}

The most common, important and consequential misconception about calorimetry is that a shower is a collection of minimum ionizing particles (mips). Already in the early days, it was realized that the signal from a high-energy electron absorbed in a sampling calorimeter
was substantially different from that of a muon that traversed this calorimeter and deposited the same energy in it as the showering electron.
This is due to the fact that the composition of the em shower changes as a function of depth, or age. 
\vskip 2mm
\begin{figure}[b!]
\epsfysize=7.5cm
\centerline{\epsffile{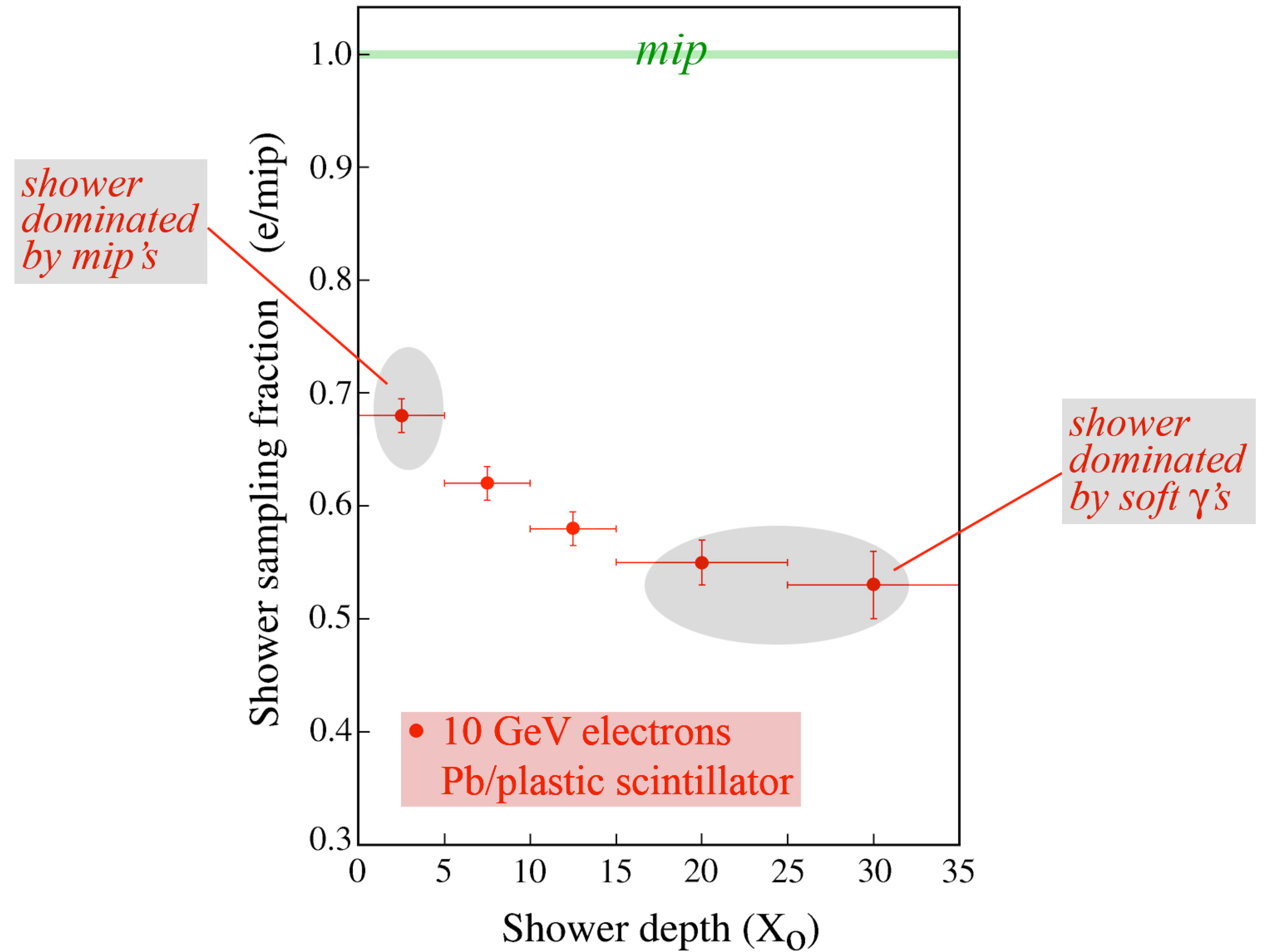}}
\caption{\footnotesize The sampling fraction changes in a developing shower. The $e/mip$ ratio is shown as a function of depth, for 10 GeV electrons in a Pb/scintillating-plastic calorimeter. Results of EGS4 calculations \cite{Wig06b}.} 
\label{emipz2}
\end{figure}
In the late stages, most of the energy is deposited by soft $\gamma$s which undergo Compton scattering or photoelectric absorption, and the sampling fraction for this shower component (\ie the fraction of the energy that contributes to the calorimeter signals) may be very different from that of the mips that dominate the early stages of the shower development. This causes major complications for the intercalibration of the different sections of a longitudinally segmented calorimeter, as is well known from the experiences of several experiments that have had to deal with this problem \cite{Atl06, Cer02}.

\subsubsection{\sl Intercalibration problems}

Figure \ref{emipz2} illustrates how the sampling fraction of a given calorimeter structure depends on the stage of the developing showers. In calorimeters consisting of high-$Z$ absorber material (\eg lead) and low-$Z$ active material (plastic, liquid argon), the sampling fraction may vary by as much as 25 - 30\% over the volume in which the absorption takes place \cite{Wig06b}.
An example of the pitfalls that this causes for calibrating a longitudinally segmented device concerns the calorimeter for the AMS-02 experiment at the International Space Station \cite{Cer02}. 
This calorimeter has eighteen independent longitudinal depth segments. Each layer consists of a lead absorber structure in which large numbers of plastic scintillating fibers are embedded, and is about $1 X_0$ thick. A minimum ionizing particle deposits 11.7 MeV upon traversing such a layer. The AMS-02 collaboration initially calibrated this calorimeter by sending muons through it and equalizing the signals from all eighteen longitudinal segments. This seems like a very good method to calibrate this detector, since all layers have exactly the same structure.
\begin{figure}[htbp]
\epsfysize=7cm
\centerline{\epsffile{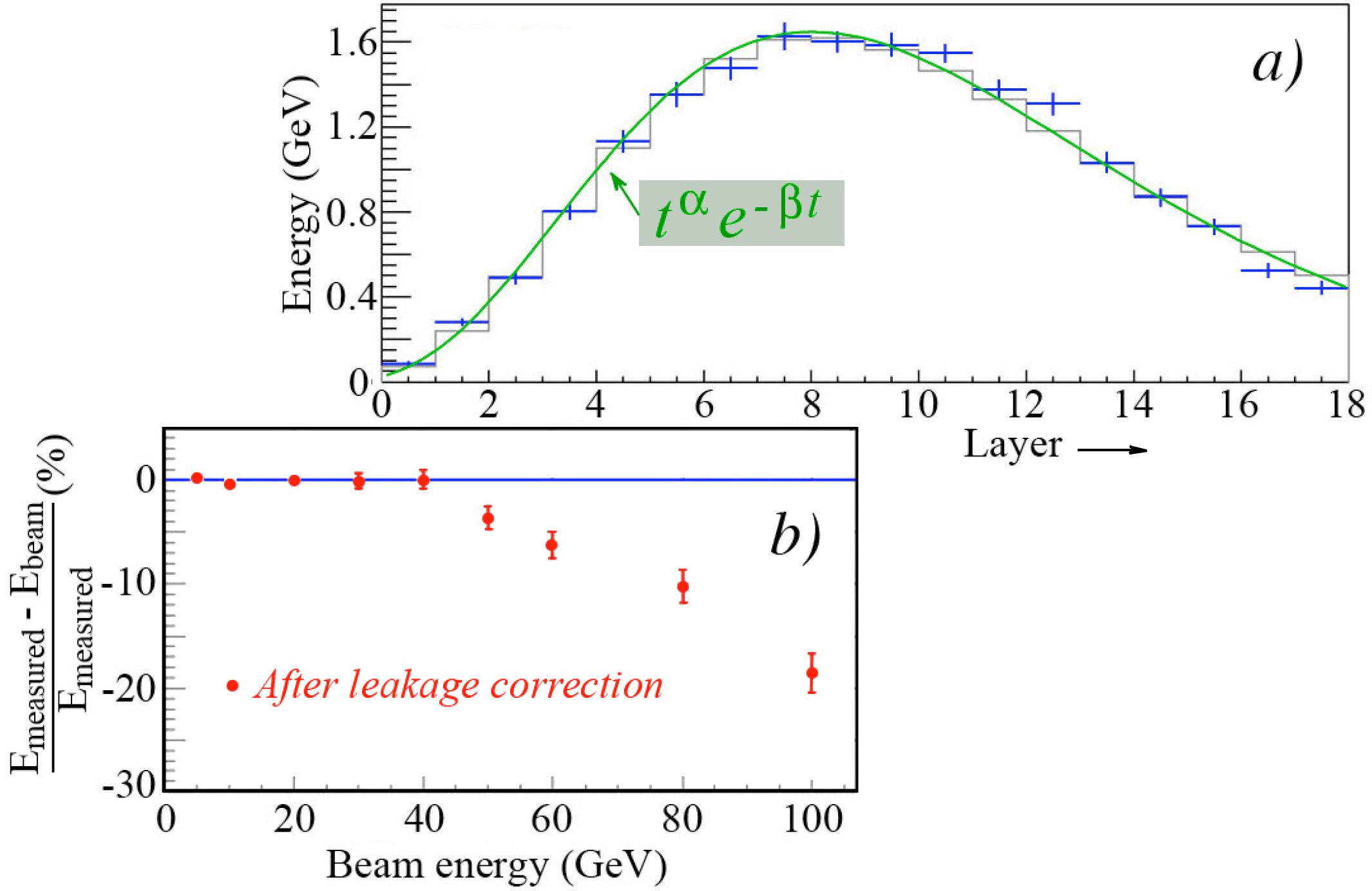}}
\caption{\footnotesize Average signals measured for 20 GeV electrons in the 18 depth segments of the AMS-02 lead/scintillating-fiber calorimeter ($a$). Average relative difference between the measured energy and the beam energy, after leakage corrections based on extrapolation of the fitted shower profile ($b$). Data from \cite{Cer02}.}
\label{ams}
\end{figure}

However, when this calorimeter module was exposed to beams of high energy electrons, it turned out to be highly non-trivial how to reconstruct the energy of these electrons.
Figure \ref{ams}a shows the average signals from 20 GeV electron showers
developing in this calorimeter. These signals were translated into energy deposits based on the described calibration.
The measured data were then fitted to a $\Gamma$-function and since the showers were
not fully contained, the average leakage was estimated by extrapolating this fit to infinity.
As shown in Figure \ref{ams}b, this procedure systematically underestimated this
leakage fraction, more so as the energy (and thus the leakage) increased. The reason for this is that a procedure in which the relationship between measured signals and the corresponding deposited energy is assumed to be the same for each depth segment will cause the energy leakage to be systematically underestimated, more so if that leakage increases. 

This very complicated problem will most definitely also affect calorimeters based on Particle Flow Analysis (PFA) \cite{Sef15}, which are all based on structures that are highly segmented, both longitudinally and laterally. The underlying problem is that the relationship between deposited energy and resulting signal is not constant throughout a developing shower. As the composition of the shower changes, so does the sampling fraction. Figure \ref{ams} provides a clear example of the problems that this may cause. 

\subsubsection{\sl Catastrophic effects}

\begin{figure}[htbp]
\epsfysize=4cm
\centerline{\epsffile{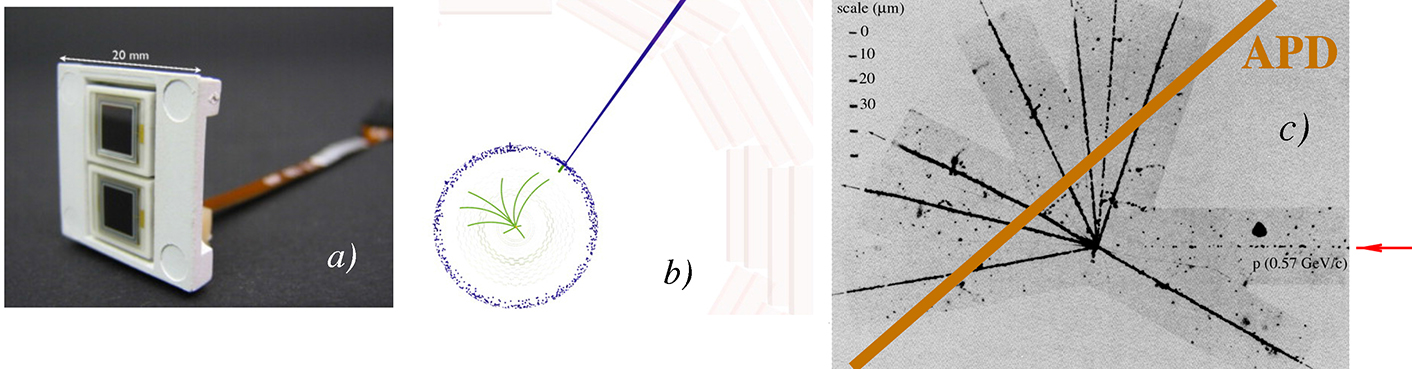}}
\caption{\footnotesize Arrangement of the APDs used to read the signals from the CMS ECAL ($a$). A ``spike'' event recorded by the CMS calorimeter system ($b$) \cite{Pet12}. A typical nuclear interaction in a developing hadron shower, induced by a 0.57 GeV/$c$ proton ($c$).} 
\label{spike}
\end{figure}

Another aspect of the misconception that a shower is a collection of mips is the fact that {\sl a single} shower particle may cause catastrophic effects for the calorimeter performance.
This is particularly true for hadron showers, and may be illustrated by a recent example taken from the CMS experiment. The CMS calorimeter system consists of a crystal based em calorimeter, followed by a brass/plastic-scintillator hadronic compartment. Each PbWO$_4$ crystal is read out by two Avalanche Photo Diodes (APDs, Figure \ref{spike}a). 
\vskip 2mm
When hadrons are sent into this calorimeter system, it produces sometimes 
events in which an anomalously large signal is recorded in one individual crystal \cite{Pet12}. Such events are referred to as ``spikes'' (Figure \ref{spike}b). Figure \ref{spike}c shows an example of a nuclear interaction that is a typical feature in hadron shower development. A proton with a momentum of 0.57 GeV/$c$ interacts with a nucleus of the absorber structure, and produces seven even lower-energy charged particles, presumably protons and/or nuclear aggregates such as $\alpha$ particles in this process. There are probably also at least as many neutrons produced in this reaction, but these do not ionize the material and are thus invisible in this figure. The charged fragments are all heavily ionizing, with typical $dE/dx$ values of 100 - 1000 times that of a mip. If such an event happens close to an APD, these
charged fragments may create a very large signal. The APDs are intended to detect scintillation photons produced in the PbWO$_4$ crystals, and the energy scale of the calorimeter signals is set by the production rate of such photons. However, the APDs  produce signals that are orders of magnitude larger when traversed by a charged particle. The densely ionizing fragments of an event such as the one shown in Figure \ref{spike}c may produce signals that are interpreted as an energy deposit of several hundred GeV inside the scintillating crystals, and this is precisely what causes these spikes. As an aside, we mention that this phenomenon should be very easily recognizable if the two APDs connected to each crystal were read out separately, since the described phenomenon would only occur in one of them. However, in order to save money, CMS had ganged them together and treated the two APDs as one readout cell.
\vskip 2mm
Another well known example of a catastrophic effect caused by a single shower particle occurs in sampling calorimeters with gaseous active media, such as proportional wire chambers that use a gas mixture containing free hydrogen atoms, \eg isobutane. Neutrons, which are abundantly produced in hadronic shower development, may elastically scatter off a hydrogen nucleus and the recoil proton may be stopped in the wire chamber. The result is a signal contribution that may be orders of magnitude larger than the signal from a mip traversing the wire chamber. Because of the extremely small sampling fraction of such calorimeters (typically $\sim \cal{O}$$(10^{-5})$, a 1 MeV energy deposit by a recoil proton is thus interpreted as a 100 GeV energy deposit in the calorimeter. This phenomenon became known as the ``Texas Tower effect'' in CDF \cite{Cih89} and necessitated a complete replacement of the forward calorimeter
system in that experiment.

\subsection{Signal (non)linearity} 

Calorimeters may be non-linear for a variety of reasons. Intercalibration of longitudinal sections, signal saturation and the energy dependence of the em shower fraction (in hadron showers) are the most common causes.
Many calorimeters are non-linear, even though their owners sometimes pretend otherwise.

A common misconception is that a calorimeter is linear if the average signals plotted versus the deposited energy can be described with a straight line.
{\bf This is incorrect}. The straight line has to extrapolate through the origin of the plot.
Signal linearity means that the average calorimeter signal is {\sl proportional} to the deposited energy, \ie the {\sl response is constant}.

\begin{figure}[htbp]
\epsfysize=6cm
\centerline{\epsffile{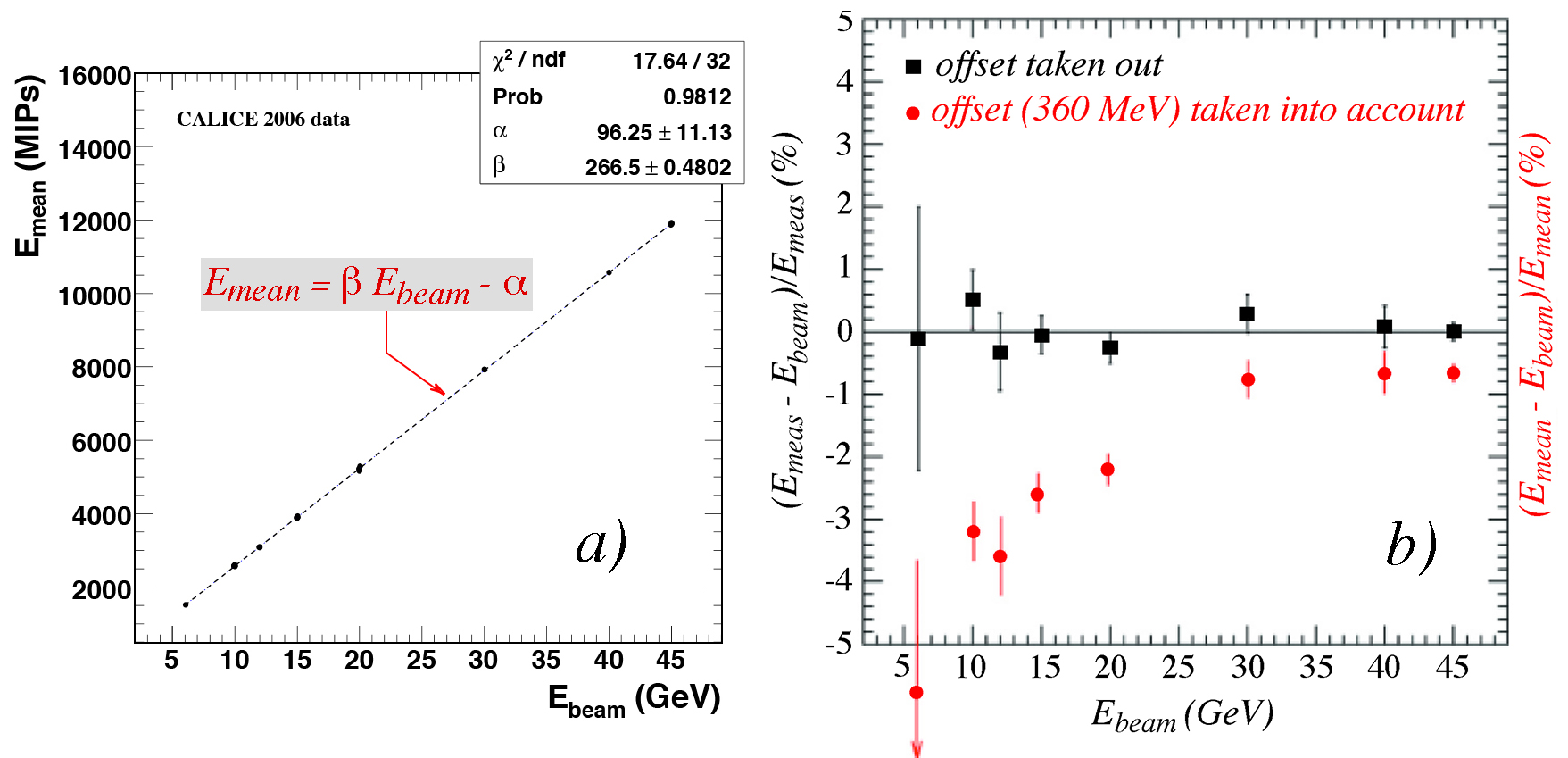}}
\caption{\footnotesize Average signal as a function of electron energy for the W/Si ECAL built by CALICE ($a$) \cite{Adl09}. Residual signals from this detector, before and after taking out a 360 MeV offset ($b$).} 
\label{nonlin}
\end{figure}

Figure \ref{nonlin} illustrates this issue. The experimental data were obtained with a W/Si em calorimeter built by CALICE \cite{Adl09}. The authors fit the measured signals with the following expression:

\begin{equation}
E_{\rm mean} = \beta ~ E_{\rm beam} - 360~{\rm MeV}
\label{sefkow1}
\end{equation}
\vskip 1mm
Then, they define
\vskip 1mm
\begin{equation}
E_{\rm meas} = E_{\rm mean} +  360~{\rm MeV}
\label{sefkow2}
\end{equation}
\vskip 1mm
and plot 
\vskip 1mm
$$(E_{\rm meas} - E_{\rm beam})/E_{\rm meas}$$
as a function of the beam energy. The result is represented by the (black) squares in Figure \ref{nonlin}b. They conclude that
``{\sl the calorimeter is linear to within approximately 1\%}.''
This is highly misleading. When the calorimeter signals they actually {\sl measured} are used to check the linearity, \ie when
\vskip 1mm
$$(E_{\rm mean} - E_{\rm beam})/E_{\rm mean}$$
is plotted as a function of the beam energy, the results, represented by the (red) full circles in Figure \ref{nonlin}b, look quite different.
We conclude from these results that the authors measured a signal non-linearity of 5\% over one decade in energy.

\subsubsection{\sl Non-linearity resulting from signal saturation}

\begin{figure}[htbp]
\epsfysize=6cm
\centerline{\epsffile{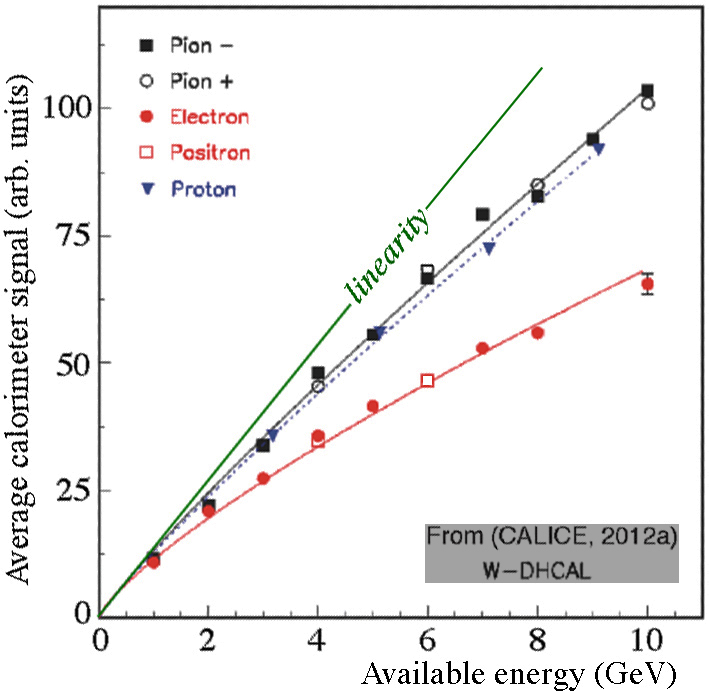}}
\caption{\footnotesize Non-linearity as a result of signal saturation.
The average signal is shown as a function of energy for electrons and hadrons in a large calorimeter based on digital readout \cite{Sef15}.}
\label{repond1}
\end{figure}

\begin{figure}[htbp]
\epsfysize=6cm
\centerline{\epsffile{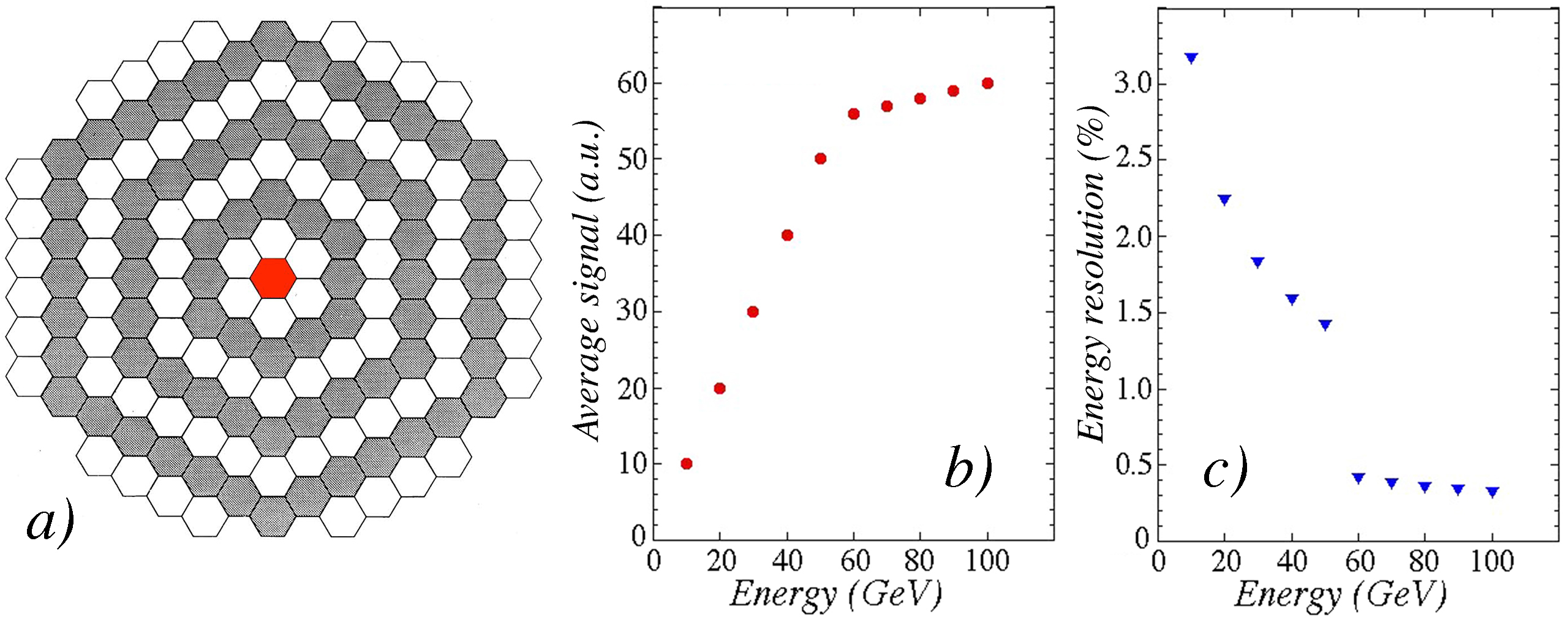}}
\caption{\footnotesize Saturation effects in one of the towers of the SPACAL calorimeter ($a$). Shown are the average signal ($b$) and the energy resolution ($c$) as a function of energy, measured when a beam of electrons is sent into this tower.}
\label{spasat}
\end{figure}

Whereas the non-linearity discussed in the previous subsection is probably the result of the intercalibration of the numerous longitudinal segments of this PFA  calorimeter, Figure \ref{repond1} shows non-linearity with a different origin. It concerns data obtained with a digital hadron calorimeter built by CALICE \cite{Sef15}. This calorimeter contains 500,000 readout cells ($1 \times 1$ cm$^2$ RPCs), which produce ``digital'' signals ("yes" or ``no'') in response to charged particles. However, this type of cell produces the same signal, regardless whether it is caused by 1, 3 or 29 shower particles. This leads to signal non-linearity, especially in em showers. Since the lateral shower profile is independent of the energy of the showering particle, and the longitudinal shower profile only varies logarithmically with that energy, the density of shower particles in the region where the energy is deposited increases almost proportionally with the  
shower energy, signal non-linearity is {\sl inevitable}. The same is true for hadron showers, albeit that the shower particle density is smaller in that case, and the non-linearity effects correspondingly less pronounced. 
\vskip 1mm
We use data from one of our own experiments to illustrate the effects of signal saturation.
The SPACAL calorimeter (Figure \ref{spasat}a) consisted of 155 hexagonal towers. Each of these towers was calibrated by sending a beam of 
40 GeV electrons into its geometric center. Typically, 95\% of the shower energy was deposited in that tower, the remaining 5\% was shared among the six neighbors. The high-voltage settings were chosen such that the maximum energy deposited in each tower during the envisaged beam tests would be well within the dynamic range of that tower. For most of the towers (except the central septet), the dynamic range was chosen to be 60 GeV. 
\vskip 1mm
When we did an energy scan with electrons in one of these non-central towers, the results shown in Figures \ref{spasat}b
and \ref{spasat}c were obtained. Up to 60 GeV, the average calorimeter signal increased proportionally with the beam energy, but above 60 GeV, a non-linearity became immediately apparent (Figure \ref{spasat}b). The signal in the targeted tower had reached its maximum value, and would from that point onward produce the same value for every event. Any increase in the total signal was due to the tails of the shower, which developed in the neighboring towers. A similar trend occurred for the energy resolution (Figure \ref{spasat}c). Beyond 60 GeV, the energy resolution suddenly improved dramatically. Again, this was a result of the fact that the signal in the targeted tower was the same for all events at these higher energies. The energy resolution was thus completely determined by event-to-event fluctuations in the energy deposited in the neighboring towers by the shower tails.

A similar situation occurred in the CALICE calorimeter of which the results are shown in Figure \ref{repond1}. And since also in this calorimeter 
an important source of fluctuations is suppressed, the energy resolution measured with it is meaningless. 

\subsubsection{\sl Non-linearity for hadron detection}

Calorimeters intended for the detection of hadron showers are typically intrinsically non-linear, as a result of the fact that the average em shower fraction depends on the energy of the showering
particle. Non-compensating calorimeters respond differently to the em and non-em shower components ($e/h \ne 1$), and the overall calorimeter response reflects 
the fact that the energy sharing between these shower components is energy dependent. 
These signal non-linearities for hadron detection are thus the result of the physics of the shower development process, they do not depend on peculiarities of the calorimeter signals, as in the examples described in the previous subsection. For that reason, hadronic signal non-linearity does in general not preclude an (on average) correct measurement of the energy of the showering particle on the basis of the observed signals. This is not necessarily true for all the non-linearities that may affect electromagnetic shower detection, such as the ones discussed in the next subsection. 
\vskip 1mm
\begin{figure}[htbp]
\epsfysize=7cm
\centerline{\epsffile{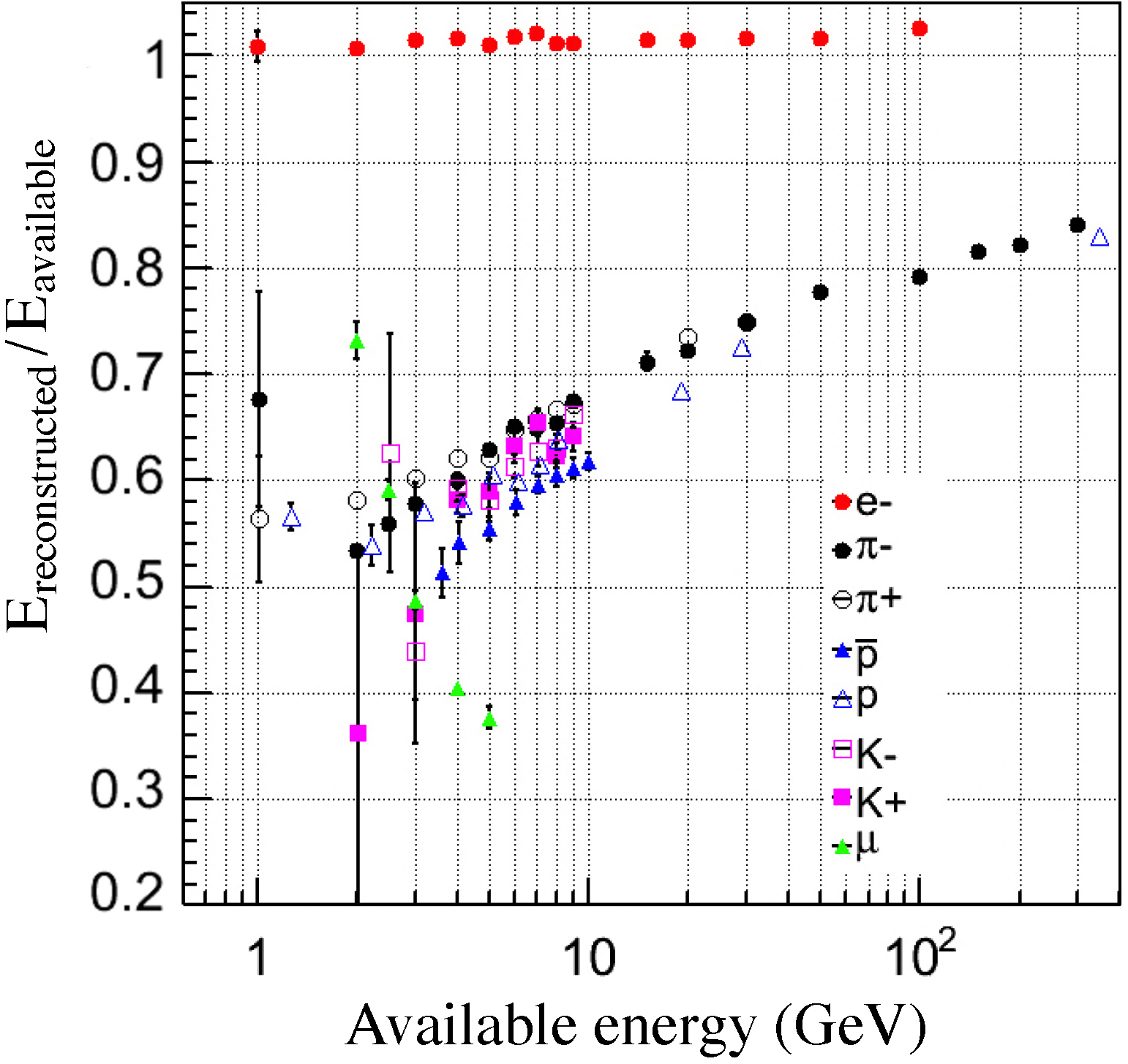}}
\caption{\small
The response to electrons and pions as a function of energy, for the CMS barrel calorimeter. The ratio of the total reconstructed energy in the calorimeter and  the available beam energy for a variety of beam particles is shown as a function of energy. The different calorimeter sections were calibrated with electrons \cite{CMS07}.}
\label{kazimcomb}
\end{figure}
A (on average) correct measurement of hadronic energy deposits is possible, provided that the em energy scale has been determined in the same way for all longitudinal calorimeter segments. In that case the hadronic response can be determined with hadron beams of different energy, using the em energy scale. An example is shown in Figure \ref{kazimcomb}. The correct hadron energy is then found by multiplying the measured energy with the inverse of the calorimeter response for that energy. The figure shows slightly different responses for different types of hadrons, but in CMS this is a secondary effect compared to the large dependence of the response on the starting point of the showers \cite{CMS07}. 

\subsubsection{\sl Signal non-linearity as a result of miscalibration}

\begin{figure}[b!]
\epsfysize=6.3cm
\centerline{\epsffile{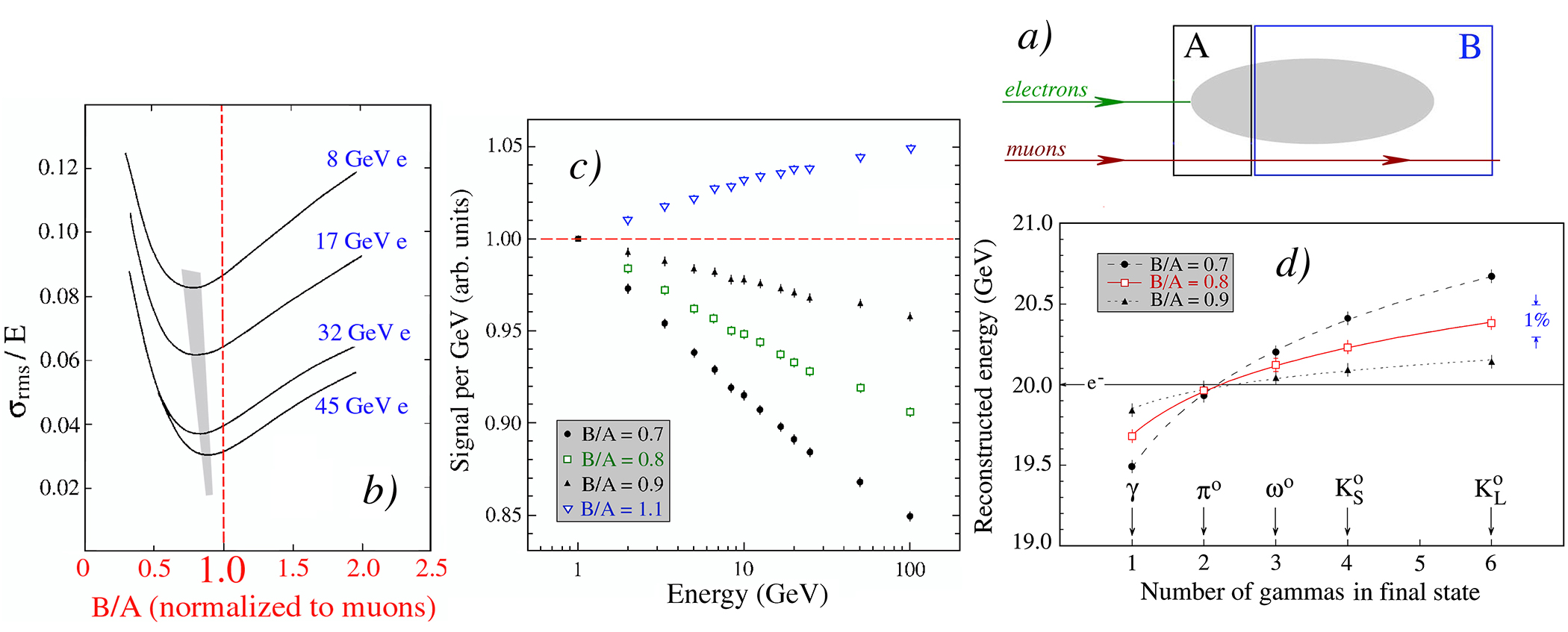}}
\caption{\footnotesize Calibration procedures and resulting effects in the HELIOS calorimeter \cite{Ake87}. The calibration constants of the two longitudinal sections ($a$) of this calorimeter ($A$ and $B$) were varied in order to minimize the total width of the electron signal distributions ($b$). The chosen value of the intercalibration constant $B/A$ leads to a signal non-linearity
($c$) for electrons. Even for the same energy (20 GeV), 
the average value of the signal depended on the number of $\gamma$s of which the showering electromagnetic object consists ($d$). See text for details.} 
\label{HEcalmethod}
\end{figure}

One of the most common reasons for signal non-linearity is the method chosen to intercalibrate the various longitudinal sections of a longitudinally segmented calorimeter. This is illustrated with the example of the HELIOS calorimeter \cite{Ake87}, discussed below. This calorimeter consisted of two longitudinal segments, with depths of $6.4 X_0$ and 4 $\lambda_{\rm int}$, respectively (Figure \ref{HEcalmethod}a). Electrons developing in this structure deposited comparable amounts of energy in each section, but the precise energy sharing depended on the energy of the showering particle. The intercalibration of the signals from the two sections was performed by minimizing the width of the {\sl total } signal distribution. Figure \ref{HEcalmethod}b shows how this width depended on the choice of the ratio of the calibration constants for the signals from both sections, $B/A$. The optimum value turned out to be different from the value for muons. The latter could simply be calculated from the composition of the two sections. 
\vskip 1mm
If we take $B/A = 1$ for muons, then the optimal value for electrons was around 0.6 - 0.9, depending on the energy. This can be understood from the fact that the sampling fraction decreased as the shower developed (Figure \ref{emipz2}). Since the sampling fraction in the first section was larger, a smaller total width was obtained when signals from that section were attributed a relatively larger weight, hence the optimal value $B/A < 1$. 
Now, if the electron energy increases, a larger fraction of that energy is deposited in the second calorimeter section. And since the signals from that section are given a relatively small weight, the result is a total signal that is smaller than if the signals from both sections had been given the same weights as mips ($B/A = 1$). In other words, the calorimeter {\sl response}
(\ie the average signal per GeV) decreases.

A calibration procedure in which the width of the total signal distribution of showers that develop in
several different calorimeter segments is minimized thus leads {\sl inevitably} to a non-linear response.  
Now one
might argue that there is in principle no reason why a  calorimeter that
is non-linear for em shower detection, although somewhat inconvenient, should be unacceptable. After
all,  all non-compensating calorimeters are intrinsically non-linear for hadron and jet detection and
one uses those too, in many experiments.

Any type of non-linearity could in principle be dealt with by means of a polynomial relationship 
between the signals $S$ and the corresponding energy $E$:
\vskip 1mm
\begin{equation}
E~=~c_0~+~c_1 S~+~c_2 S^2~+~c_3 S^3~+~...  
\label{polycalib}
\end{equation}
and the fact that other constants than $c_1$ have a non-zero value might be a small price to pay for
improving energy resolution. 

This line of reasoning is, however, crucially flawed \cite{Mehmet}. The non-linearity introduced by this weighting
scheme implies {\sl by definition} that a high-energy $\pi^0$, decaying into two unresolved $\gamma$s
produces, on average, a larger signal in this calorimeter than an electron, or one photon, of the same
energy. An $\omega^0$ resonance decaying into three unresolved $\gamma$s produces an even larger signal,
and an energetic $K^0$ decaying into $\pi^0 \pi^0$, or even $\pi^0 \pi^0 \pi^0$ tops them all (Figure \ref{HEcalmethod}d).
By introducing a signal non-linearity, the calorimeter response is made dependent on such
differences. And since, in practice, the calorimeter information does not always allow one to tell
whether the signal was caused by one, two, three or even more $\gamma$s, the systematic differences in the
average calorimeter response for those cases are an {\sl integral part of the energy resolution}.
Interpreting the width of the signal distribution measured for single electrons from a test beam as the
em energy resolution is thus incorrect.
\vskip 2mm

The approach chosen in this case (minimization of the width of the total signal distribution) is only one of several different methods described in the literature for intercalibrating
the different sections of a longitudinally segmented calorimeter. Other methods aim to achieve 
\begin{enumerate}
\item Correct energy reconstruction of pions penetrating the em compartment without starting a shower, or
\item Hadronic signal linearity, or
\item Independence of hadron response on starting point shower, or
\item Equal response to electrons and pions.
\end{enumerate}
Each of these approaches introduces specific additional problems \cite{OUP2}.
Intercalibrating the different sections of a longitudinally segmented calorimeter system is in practice one of the
most daunting tasks when commissioning a detector, and it is {\sl fundamentally impossible} to achieve a result in which
the signals measured in the different sections can be correctly translated into deposited energy. This is even true for
compensating calorimeters. The combination of the energy dependence of the shower profiles, combined with the
depth dependence of the sampling fraction are responsible for this problem.
     
The best way to intercalibrate the 
different sections of a longitudinally segmented calorimeter system is by using 
the same particles for all individual sections. If these particles develop showers, then they can only be
used to calibrate sections in which these showers are completely contained.
Only in this way is the relationship between the deposited shower energy (in GeV) and the charge
(in picoCoulombs) generated as a result established unambiguously.
We have referred to this as the $B/A = 1$ method.
The use of a beam of muons to intercalibrate the eighteen segments of the AMS-02 electromagnetic calorimeter (Figure \ref{ams}) definitely qualifies as a viable method in this respect.
The mistake made in that case did not concern the calibration method itself, but the interpretation of the results.

\subsection{Energy resolution}

A common mistake with regards to energy resolution has to do with its very definition. The energy resolution is the precision with which the energy of an unknown object can be determined from the signals it produces in the calorimeter. Typically, this resolution is determined as the relative width of the signal distribution measured for a beam of mono-energetic particles from an accelerator. However, this is only correct if the average value of that measured signal distribution corresponds indeed to the correct energy of these particles.
Response non-linearities tend to invalidate that assumption, as illustrated by the example shown in Figure \ref{HEcalmethod}d. 

Often, the measured signal distributions exhibit non-Gaussian tails. In that case, one should quote the $\sigma_{\rm rms}$ value as the energy resolution. However, some authors 
use another variable, in order to make the results less dependent on the tails of the signal distributions they measure, and thus look better. This variable, called $\rm rms_{90}$, is defined as the root-mean-square of the energies located in the smallest range of reconstructed energies which contains 90\% of the total event sample.
For the record, it should be pointed out that for a perfectly Gaussian distribution, this variable gives a 21\% smaller value than the true $\sigma_{\rm rms}$ (\ie $\sigma_{\rm fit}$).
Of course, one is free to define variables as one likes. However, one should then not use the term ``energy resolution'' for the results obtained in this way, and compare results obtained in terms of $\rm rms_{90}$ with genuine energy resolutions from calorimeters with Gaussian response functions \cite{Tho09}. This misleading practice is generally followed by the proponents of PFA.

\vskip 2mm
Another widespread misconception concerns the way in which the energy resolution of a calorimeter is quoted. Frequently, the relative energy resolution ($\sigma/E$) of a particular calorimeter is expressed as $x\%/\sqrt{E}$. However, this is rarely a correct description of reality, since in practice other factors, which are not governed by Poisson statistics, contribute to the energy resolution, and such factors often dominate the performance, especially at the low and high ends of the energy spectrum for which the detector is intended.
\vskip 1mm
\begin{figure}[htbp]
\epsfysize=7.5cm
\centerline{\epsffile{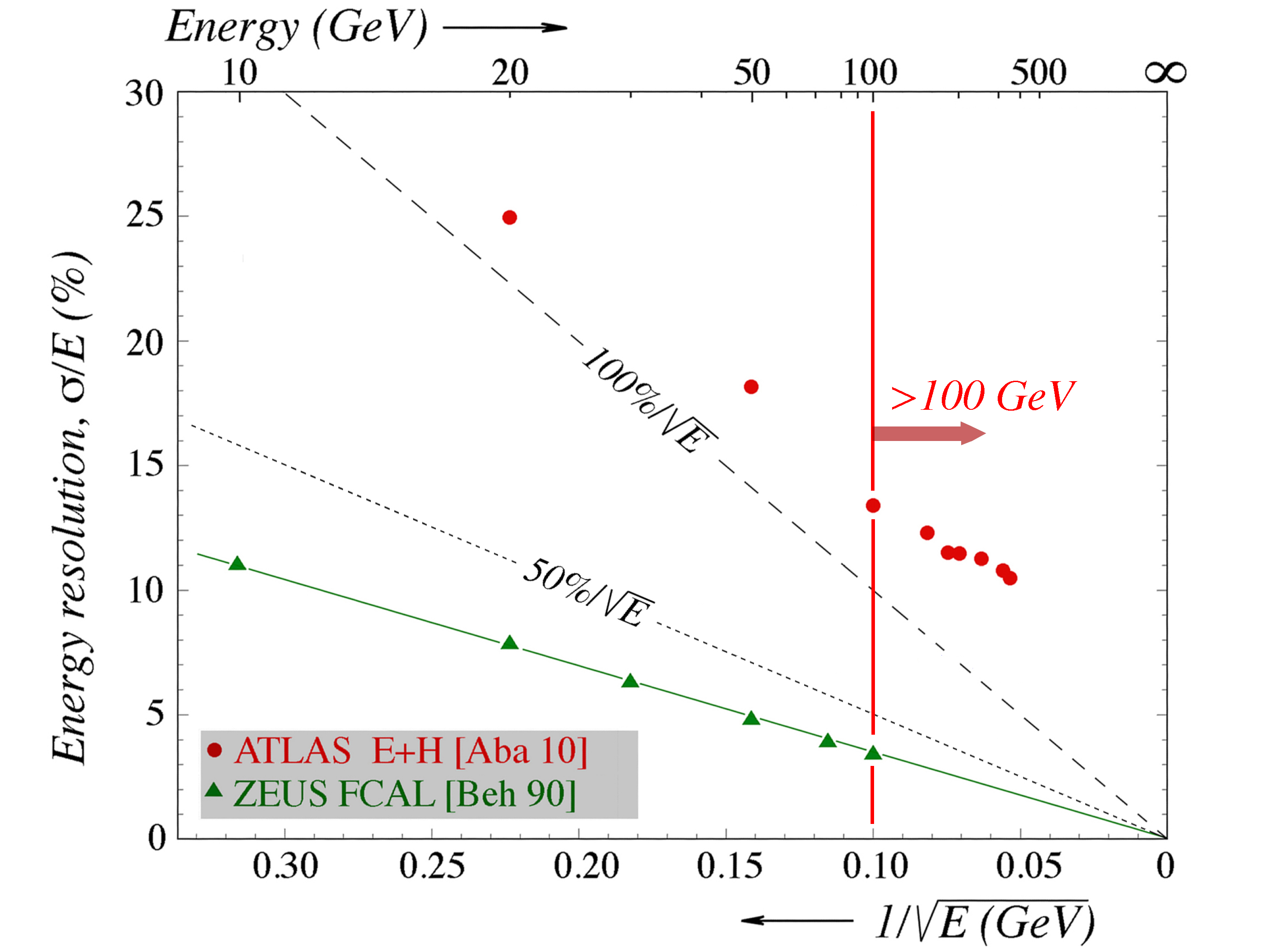}}
\caption{\footnotesize Hadronic energy resolutions of the ZEUS \cite{Beh90} and ATLAS \cite{Aba10} calorimeters.} 
\label{zatlas}
\end{figure}
As an example, Figure \ref{zatlas} shows the hadronic energy resolutions of the ZEUS and the ATLAS calorimeters. The experimental data points are plotted on a scale that is linear in $E^{-1/2}$ and runs from right to left. Scaling with $1/\sqrt{E}$ implies that the data points should be located on a straight line through the bottom right corner in this plot. This is indeed the case for the compensating ZEUS calorimeter, for which the resolution is quoted as $35\%/\sqrt{E}$. However, the resolution of ATLAS does not at all scale with $1/\sqrt{E}$. As a matter of fact, the data points are at all energies located well above the $100\%/\sqrt{E}$ line in this plot, and for energies larger than 100 GeV, the resolution is more than a factor of four worse than for ZEUS. Yet, in talks about ATLAS, the hadronic energy resolution is often 
quoted as $0.6 - 0.7/\sqrt{E}$.

\vskip 2mm
Another mistake that is not uncommon concerns the extrapolation of measurement results far beyond their region of validity, We mention two examples. 

The HELIOS
Collaboration measured a resolution $\sigma/E = 1.9\%$ for 3.2 TeV $^{16}$O
ions \cite{Ake87}, and the WA80 Collaboration, which also operated a uranium/scintillator
calorimeter, found a resolution of 1.7\% for 6.4 TeV $^{32}$S ions \cite{You89}.
One should realize, however, that in these cases a convolution of either 16 or 32 independent 
200 GeV nucleon showers was measured. Hence, strictly speaking, these results only say something
about the precision of the energy measurement for a 200 GeV nucleon shower. If sixteen signals from such
showers are convolved, then the resulting signal has a resolution $\sigma/E$ that is four (=
$\sqrt{16}$) times smaller than the resolution for the individual signals from 200 GeV nucleons. In
other words, if the resolution for 200 GeV protons (or neutrons) was 7.6\%, then a resolution of
1.9\% should be expected for $^{16}$O ions with an energy of 3.2 TeV.
The measured resolution for heavy ions at multi-TeV energies is thus by no means indicative for the
resolution that may be expected for the detection of single hadrons or jets carrying such energies.

A similar statement should be made concerning the ``determination'' of the energy resolution for
high-energy em shower detection in liquid xenon, based on convolving the signals from
large numbers of low-energy electrons (100 keV) recorded in a small cell \cite{Seg92}.
Also in this case, the measurements only revealed something about the energy resolution for the
detection of these low-energy electrons. In a high-energy em shower, a variety of new
effects, absent or negligible in the case of these electrons, affect the signals and their
fluctuations. As an example of such effects, we mention the fact that the (174 nm) shower light is
produced in a large detector volume. Light attenuation, \eg\ through self-absorption and
shower leakage, are the likely consequences of this.

These examples illustrate that, in general, measurements made for low-energy particles cannot be
used to determine the high-energy calorimeter performance.  
\vskip 2mm

Finally, we want to point out that often times a good energy resolution is only part of the requirements for obtaining the desired physics sensitivity.
As an example, we mention the Higgs boson, discovered in 2012 by two experiments at the Large Hadron Collider through its decay mode 
$H^0 \rightarrow  \gamma \gamma$ \cite{Aad12,Chat12}.
The invariant mass of a particle decaying into two $\gamma$s is given by
\vskip 1mm
\begin{equation}
M~=~\sqrt{2 E_1 E_2 (1 - \cos{\theta_{12}})}
\label{invmas2g}
\end{equation}
The precision with which the mass can be measured is thus not only determined by the energy resolution,
\ie the measurement uncertainty on the $\gamma$ energies $E_1$ and $E_2$, but also by the 
relative uncertainty on the angle ($\theta_{12}$) between the directions of these $\gamma$s.
A good localization of the $\gamma$s is thus very important to identify the parent particle.
While CMS emphasized excellent energy resolution for em showers in its design of the experiment, at the expense of degraded
hadronic performance, ATLAS concentrated its efforts also on the localization issue. As a result, the mass resolution for the
Higgs bosons turned out to be very similar in both experiments.
  
\subsection{Effects of non-compensation}

Almost all calorimeters that are operating in large storage ring experiments are non-compensating. This means that the responses (\ie the average signal per unit deposited energy) to the em and non-em components of hadron showers are not the same in these calorimeters ($e/h  \ne 1.0$). The consequences of this feature are a source of several misconceptions. Often, an additional {\sl constant term} in the hadronic energy resolution is considered the main, if not the only, consequence of non-compensation. This is a misconception at several levels. Not only is non-compensation the cause of a number of other serious problems, but the effect on the hadronic energy resolution is by no means independent of energy, as suggested by the concept of an additional {\sl constant} term.
\vskip 1mm
\begin{figure}[htbp]
\epsfysize=6cm
\centerline{\epsffile{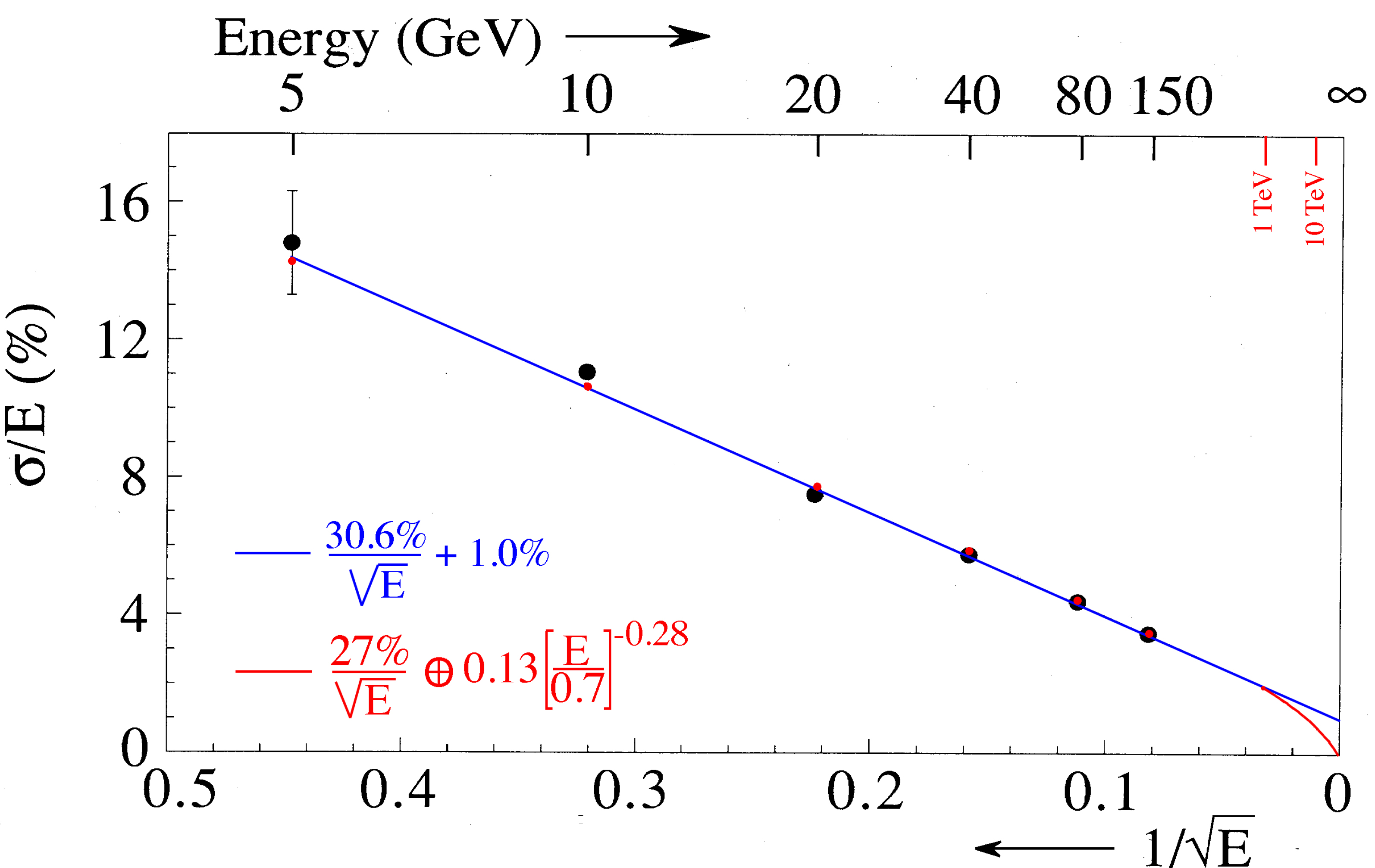}}
\caption{\footnotesize Experimental data points for the hadronic energy resolution of the SPACAL lead/plastic-fiber calorimeter 
\cite{Aco91c}, together with two different fits of these data points. The differences between these two fits only become noticeable at 
very high energies, beyond the range where this instrument was tested. } 
\label{effects_noncomp}
\end{figure}

The incorrectness of this notion is illustrated by the fact that the hadronic energy resolution of non-compensating calorimeters is not only considerably worse compared to compensating ones at high energies, but also at low energies. The resolution of the best hadron calorimeters, such as the one used for the ZEUS experiment \cite{Beh90}, is $\sim 30\%/\sqrt{E}$, \ie
$\sigma/E \sim 10\%$ at 10 GeV (Figure \ref{zatlas}). Adding a constant term of 5\% would increase this resolution to $\sim 11\%$. However, the energy resolution of non-compensating calorimeters is typically 
two to three times larger at this energy. The effects of non-compensation are thus by no means limited to high energy, where a constant term tends to dominate the contributions that are determined by Poisson fluctuations.

The correct way of incorporating the effects of non-compensation on the hadronic energy resolution is given by Equation \ref{lognoncomp}, and illustrated in Figure \ref{effects_noncomp}.
\vskip 1mm
\begin{equation}
{\sigma\over E}~=~{a_1\over \sqrt{E}} \oplus a_2 \biggl[ \biggl({E\over E_0}\biggr)^{l-1}\biggr]
\label{lognoncomp}
\end{equation}
The effects are described by an energy dependent term, added in quadrature to the scaling term that accounts for the Poisson fluctuations ($a_1 = 27\%$ in this example). 
The coefficient of this non-compensation term, $a_2 = 0.13$ in this example, is determined by the degree of non-compensation: 
$a_2 = |1 - h/e|$, and $l \sim 0.72$ \cite{deg07}.

Figure \ref{effects_noncomp} also shows that the correct description of the hadronic energy resolution yields in practice almost identical results as an expression in which a constant term ($1\%$) is added linearly to a scaling term ($30\%/\sqrt{E}$). There are several examples in the literature in which such an expression is used to describe the hadronic energy resolution. However, the linear addition of two terms suggests complete correlation between the effects described by these terms, which is nonsense in this situation. Figure \ref{effects_noncomp} shows that one has to go to very high energies, beyond the reach of the current generation of available test beams, to see a significant difference between the two mentioned expressions.
However, only one of these expressions (the red one) is correct, and indicates that the effects of non-compensation on the hadronic energy resolution are indeed energy dependent.
\vskip 2mm

However, a hadronic energy resolution that deviates from $E^{-1/2}$ scaling is not the only consequence of non-compensation.
Among the other effects, we mention
\begin{itemize}
\item Hadronic signal non-linearity (see Section 2.2.2). This is a result of the fact that the average em fraction of hadron showers, $\langle f_{\rm em} \rangle$, increases with energy.
\item Non-Gaussian response functions. This is a consequence of the fact that the distribution of $f_{\rm em}$ is not Gaussian, but asymmetric, favoring large values
(Figure \ref{spafem}). \vskip 1mm
\begin{figure}[htbp]
\epsfysize=5.5cm
\centerline{\epsffile{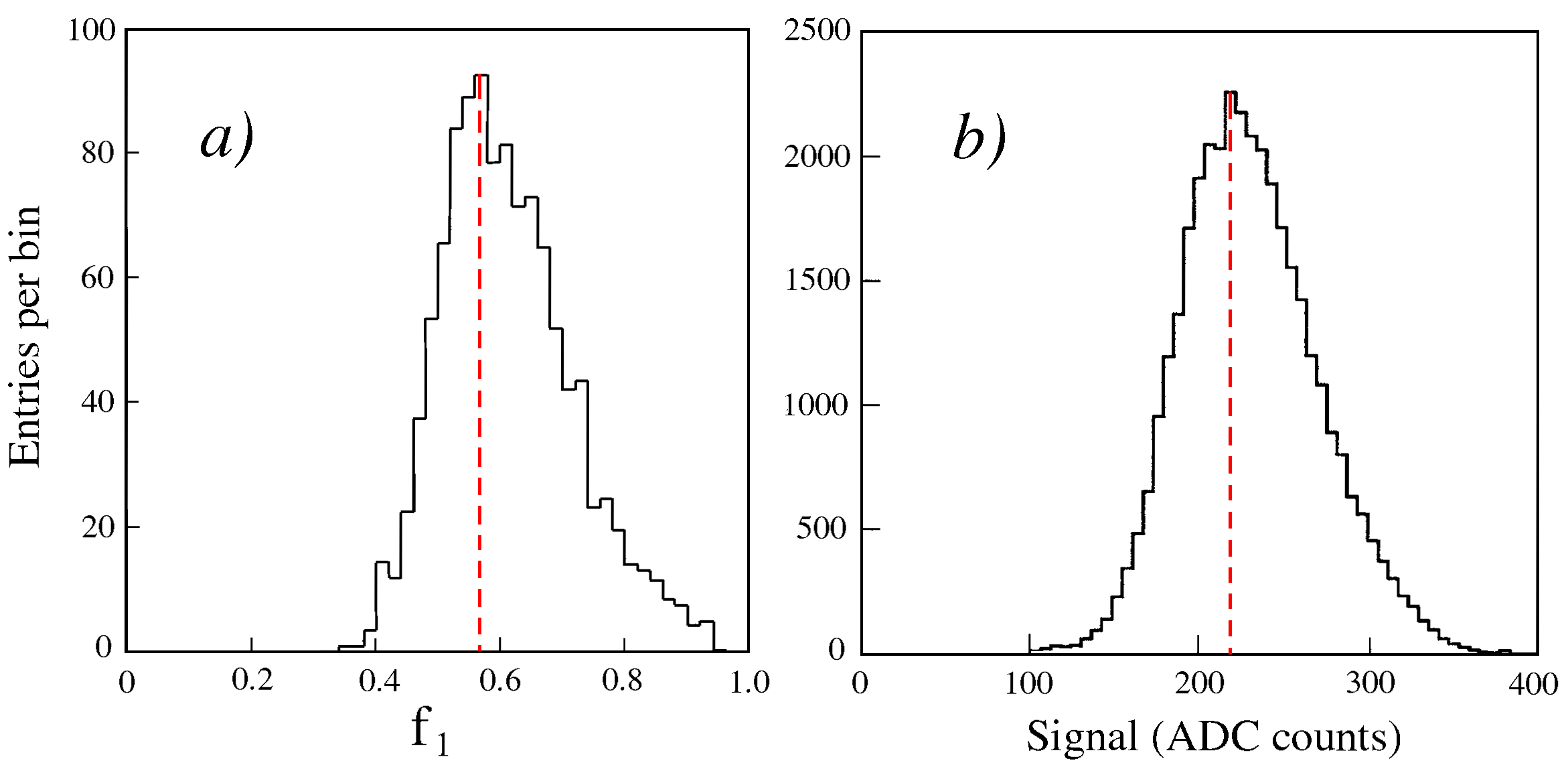}}
\caption{\small
The distribution of the fraction of the energy of 150 GeV $\pi^-$ showers contained in the em
shower core ($a$) \protect\cite{Aco92b} and the signal
distribution for 300 GeV $\pi^-$ showers in a non-compensating calorimeter ($b$)
\protect\cite{Akc98}.}
\label{spafem}
\end{figure}
This may cause problems, such as trigger biases. For
example, if one uses the calorimeter signals to select events with a minimum (missing)
transverse energy from a steeply falling distribution, then the event sample is likely to
be strongly dominated by events in which the actual value of this energy was smaller
than the trigger level, but in which upward fluctuations pushed it beyond that level. An
asymmetric response function makes it very difficult to deal with this problem in
a correct way.
\item Different response functions for different hadrons (protons, pions, kaons) of the same energy. This is the topic of the next subsection.
\end{itemize}

\subsection{Particle dependence of the calorimeter response}

The absorption of different types of hadrons in a calorimeter may differ in very fundamental ways, as a result of 
applicable conservation rules. For example, in interactions induced by a proton or neutron, conservation of baryon number has important consequences.
The same is true for strangeness conservation in the absorption of kaons.
This has implications for the way in which the shower
develops. For example, in the first interaction of a proton, the leading particle has to be a baryon. This precludes the production
of an energetic $\pi^0$ which carries away most of the proton's energy. Similar considerations apply in the absorption of 
strange particles. On the other hand, in pion-induced showers it is not at all uncommon that most of the energy carried by the
incoming particle is transferred to a $\pi^0$. The resulting shower is in that case almost completely electromagnetic. This phenomenon
is the reason for the asymmetric distributions from Figure \ref{spafem}. \vskip 1mm
\begin{figure}[htbp]
\epsfysize=6.5cm
\centerline{\epsffile{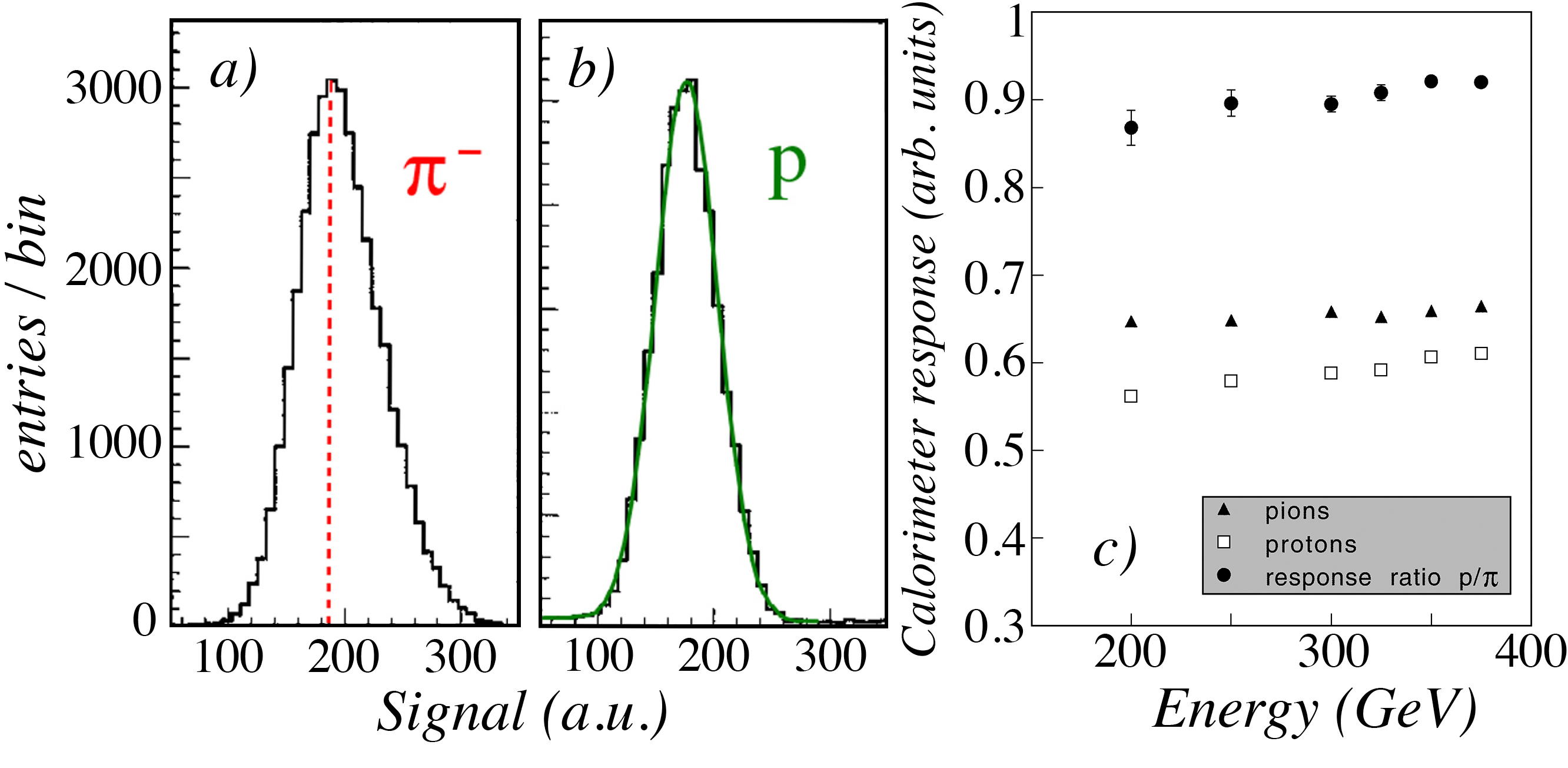}}
\caption{\small
Signal distributions for 300 GeV pions ($a$) and protons ($b$) in the CMS forward calorimeter. Average signals per GeV for protons and 
pions as well as the ratio of these response values in this detector, as a function of energy ($c$) \cite{Akc98}.}
\label{pdif}
\end{figure}

Experimental studies have confirmed these effects. Figure \ref{pdif} shows the signal distributions measured for 300 GeV pions ($a$) and protons ($b$), respectively.
The signal distribution for protons is much more symmetric, as indicated by the Gaussian fit. This is because the em component of proton-induced showers is typically populated by $\pi^0$s that share the energy contained in this component more evenly than in pion-induced showers.The figure also shows that the rms width of the proton 
signal distribution is significantly smaller (by $\sim 20\%$) than for the pions. Figure \ref{pdif}c shows that the average signal per GeV deposited energy is smaller for the protons than for the pions, by about 10\%. This is also a consequence of the limitations on $\pi^0$ production that affect the proton signals in this non-compensating calorimeter ($e/h > 1$). So while the response to protons is smaller in this calorimeter, the energy resolution is better. Similar effects are expected to play a role for the detection of kaons, where $\pi^0$ production is limited as a result of strangeness conservation in the shower development.
\vskip 2mm
Whereas the phenomena discussed above are the result of differences in the em shower component, which lead to differences in the response functions of the calorimeter to baryons, pions and kaons, other effects may also cause significant differences that at first sight might be unexpected. As an example, we mention the differences between electron and photon detection in a calorimeter. These are important, since the em performance is typically experimentally studied with electron beams, whereas photon detection may be the most important goal\footnote{As an example, we mention studies of the decay of Higgs bosons through the process $H^0 \rightarrow \gamma \gamma$ at the LHC experiments CMS and ATLAS.}. Showers initiated by high-energy photons and electrons are quite different in the early stage of the absorption process, before the shower maximum \cite{Mehmet}.
\begin{enumerate}
\item Photon-induced showers deposit their energy, on average, deeper inside the absorbing structure
than do em showers induced by charged particles of the same energy. The response differences in Figure \ref{HEcalmethod}d are the result of this.
\item The fluctuations in the amount of energy deposited in a given slab of material are larger for
showers induced by photons than for showers induced by $e^+$ or $e^-$.  
\end{enumerate}
The first effect results from the fact that the photons travel a certain distance (9/7 $X_0$, on average) in
the absorbing structure before they start losing energy, while electrons and positrons start losing
energy immediately upon their entry. Moreover, the starting point of the photon-induced showers
fluctuates from event to event, which leads to the second effect.
\vskip 1mm
\begin{figure}[tbp]
\epsfysize=6cm
\centerline{\epsffile{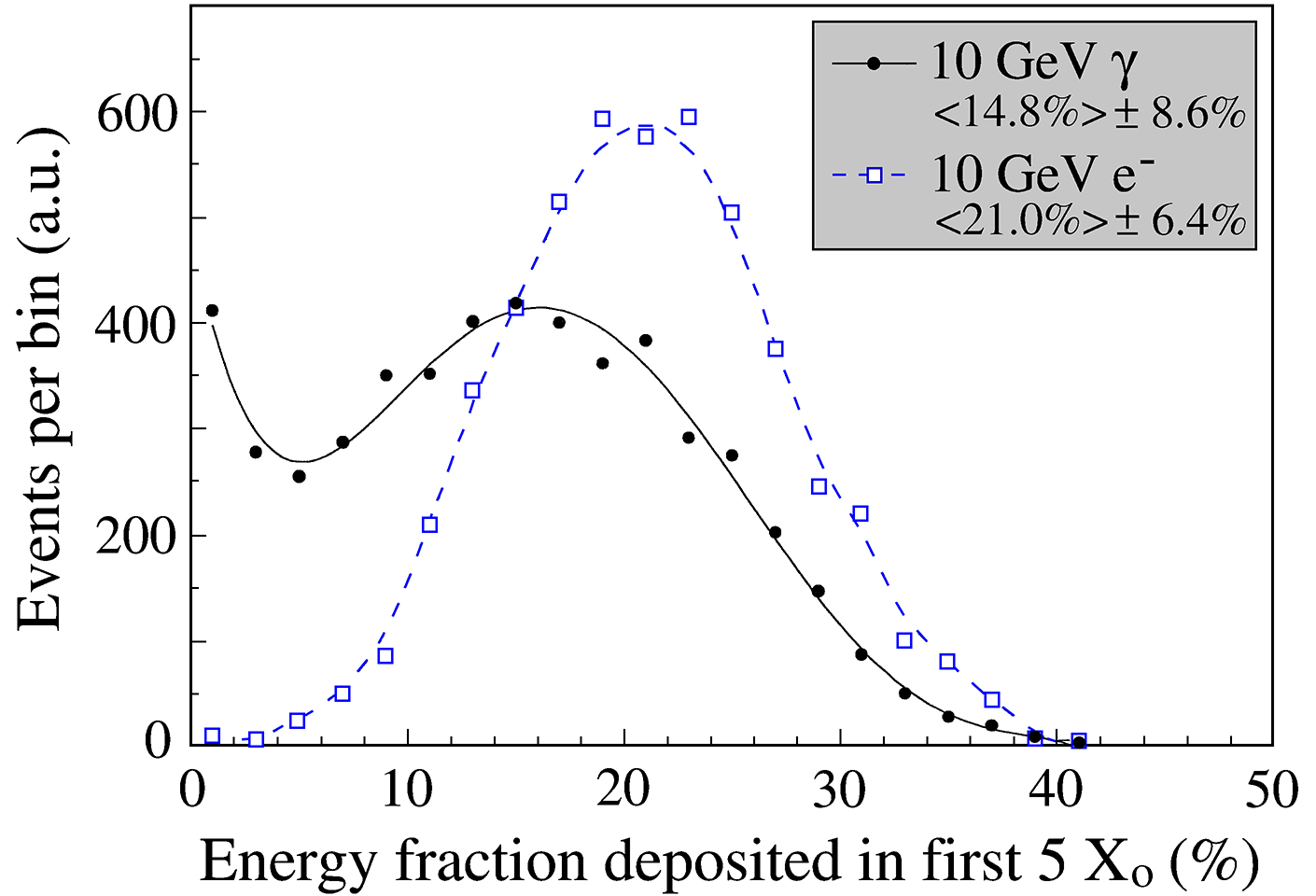}}
\caption{\small
Distribution of the energy fraction deposited in the first five radiation lengths by 10 GeV
electrons and $\gamma$s showering in lead. Results of EGS4 simulations \cite{Mehmet}.}
\label{Pb5X0}
\end{figure}

These effects are illustrated in Figure \ref{Pb5X0}, which shows the distribution of the
energy deposited by 10 GeV electrons and 10 GeV photons in a $5 X_0$ (2.8 cm) thick
slab of lead. On average, the electrons deposit more energy in this material than the photons
(2.10 GeV \vs 1.48 GeV). However, the {\sl fluctuations} in the energy deposited by the photons are
clearly larger than those in the energy deposited by the electrons (0.86 GeV \vs 0.64 GeV).
The distribution for the photon showers exhibits an excess near zero, which is the result of
photons penetrating (almost) the entire slab without interacting. The ``punch-thru'' probability
for a high-energy $\gamma$ is in this example $\exp{(-35/9)} \approx 2\%$.   

The different effects of dead material installed in front of the calorimeter on electrons/positrons and $\gamma$s
is relevant for experiments such as ATLAS, where the electromagnetic calorimeter is ``hidden'' in a 
cryostat, although the fact that this cryostat is made of aluminium makes the effects less dramatic than 
suggested in Figure \ref{Pb5X0}. Another consequence of the differences between electron and $\gamma$ induced showers is the fact that the very complicated 
calibration scheme that was developed for electrons showering in the three longitudinal segments of the ATLAS ECAL \cite{Atl06}
is {\sl not} necessarily the optimal solution for $\gamma$ detection in this calorimeter. 

\subsection{The perceived benefits of longitudinal segmentation}

There is a deeply rooted belief that calorimeter systems for high-energy collider experiments should be longitudinally subdivided into several sections.
As a minimum, one will usually want to have an electromagnetic and a hadronic section. A major reason for this belief is that such a subdivision is needed
for recognizing em showers, and thus identify electrons and $\gamma$s entering the calorimeter. 

This is a myth. It has been demonstrated repeatedly that there are several ways to identify em showers in longitudinally {\sl unsegmented} calorimeters.
For example, the DREAM Collaboration has demonstrated four different methods that can be used to achieve this \cite{pid}. These methods are based on
\begin{enumerate}
\item The measured lateral shower profile,
\item A comparison between the scintillation and \v{C}erenkov signals produced by the developing shower,
\item The time structure of the signals, and in particular 
the starting time of the signals with respect to the signal produced in an upstream detector, or
the pulse width
\end{enumerate}

\begin{figure}[htbp]
\epsfysize=9cm
\centerline{\epsffile{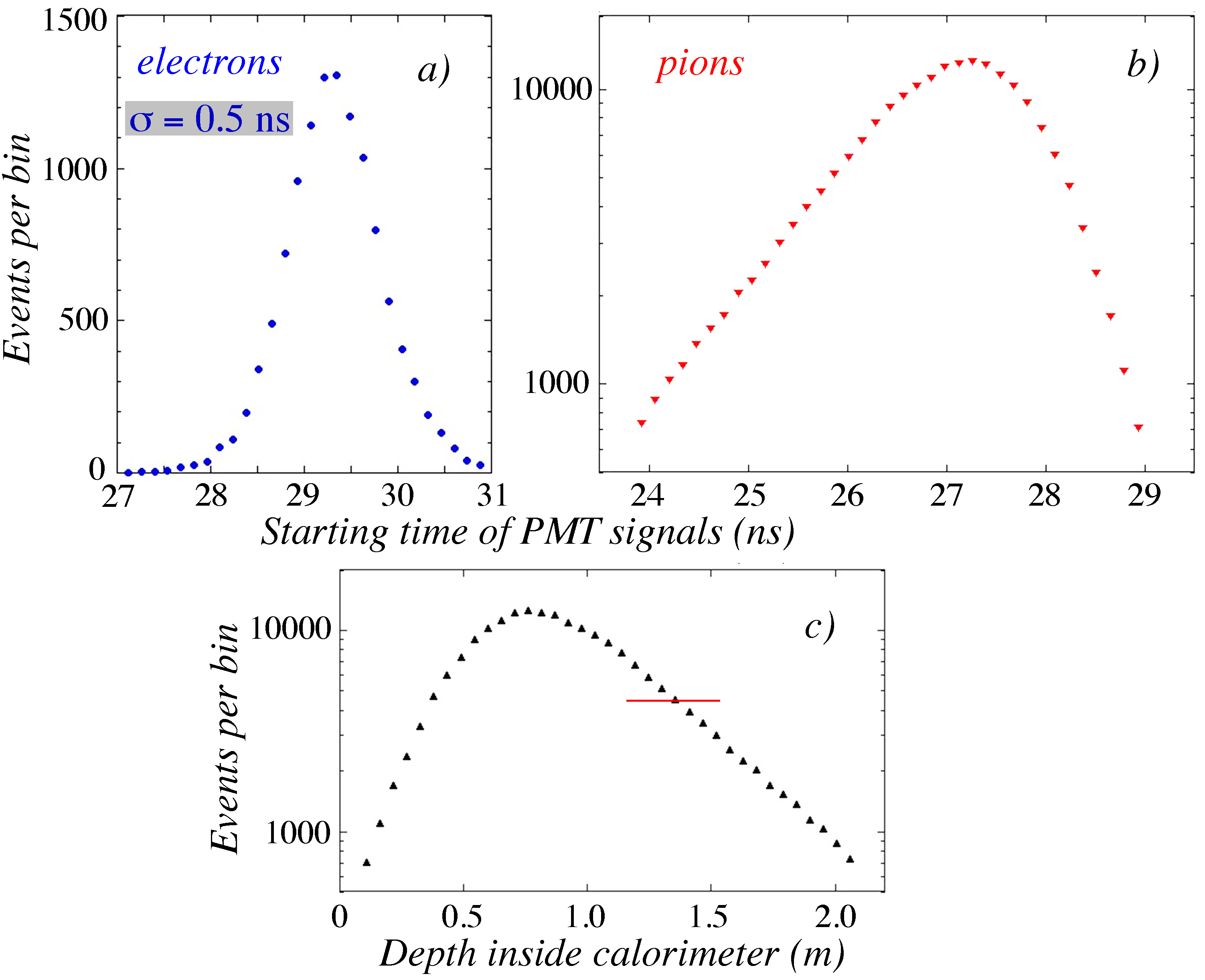}}
\caption{\footnotesize The measured distribution of the starting time of the calorimeter's scintillation signals produced by 60 GeV electrons ($a$) and 60 GeV pions ($b$). This time is measured with respect to the moment the beam particle traversed a trigger counter  installed upstream of the calorimeter. These data were also used to determine the distribution of the average depth at which the light was produced in the hadron showers ($c$).}
\label{timing}
\end{figure}

Figure \ref{timing} illustrates one of these methods, which is based on the starting time of the calorimeter signals, measured with respect to the signal
produced by an upstream detector.
This method is based on the fact that the light in the optical fibers travels at a lower speed than the particles that generate this light.
The deeper inside the calorimeter the light is produced, the earlier the calorimeter signal starts. For the polystyrene fibers, the effect amounted to 2.55 ns/m. For the tested calorimeter, this led to a longitudinal position resolution of $\sim 20$ cm.

Figure \ref{timing} shows the measured distribution of the starting time of the signals from 60 GeV $e^-$ (Figure \ref{timing}a) and $\pi^-$ (Figure \ref{timing}b). This pion distribution peaked $\sim 1.5$ ns earlier than that of the electrons, which means that the light was, on average, produced 60 cm deeper inside the calorimeter. The distribution is also asymmetric, it
has an exponential tail towards early starting times, \ie light production deep inside the calorimeter. This signal distribution was also used to reconstruct the average depth at which the light was produced for individual pion showers. The result, depicted in Figure \ref{timing}c, essentially shows the longitudinal profile of the 60 GeV pion showers in this calorimeter.

It was shown in this paper that by combining all the available methods, which in several different ways exploited complementary information about the events, 
the longitudinally unsegmented RD52 fiber calorimeter could be used to identify electrons with a
very high degree of accuracy. Using the time structure of the signals, the lateral shower profile and a comparison of the \v{C}erenkov and scintillation signals, more than 99\% of the electrons entering the detector were correctly identified with criteria that ruled out almost all hadronic particles as electron candidates. 

However, good electron/pion separation can already be achieved with much less sophisticated methods.
Especially in beam tests, a very simple preshower detector (PSD), placed in front of the calorimeter, may do an adequate job.
Such a device may consist of a plate of lead, 1 cm (1.9 $X_0$, 0.06 \lam) 
thick, followed by a sheet of plastic scintillator. When a beam consisting of a mixture of 
high-energy electrons
and pions is sent through this device, almost all pions (96\%) traverse it without strongly 
interacting. These pions produce a minimum ionizing peak in the scintillator. On the other hand,
the electrons lose a considerable fraction of their energy by radiating large numbers of 
bremsstrahlung photons. Some of these photons convert into $e^+e^-$ pairs in the PSD and
thus contribute to the scintillation signals produced by this device.
\vskip 1mm
\begin{figure}[htbp]
\epsfysize=7cm
\centerline{\epsffile{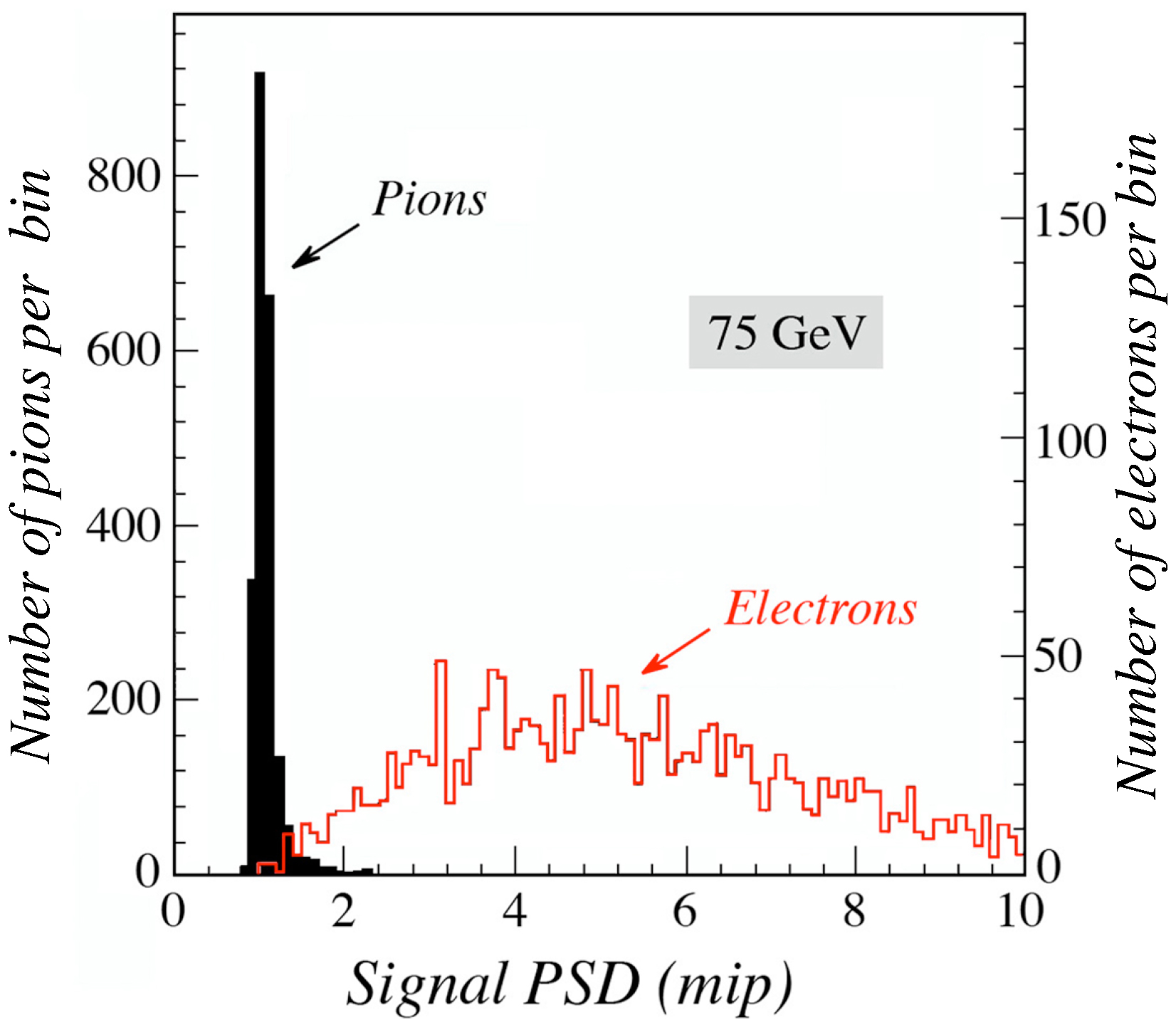}}
\caption{\small
Signal distributions for 75 GeV pions and electrons in a preshower detector used in beam tests of
CDF calorimeters.}
\label{cdfpsd}
\end{figure}

The result is a very clear separation between electrons and pions. Figure \ref{cdfpsd} shows the
signal distributions for 75 GeV electrons and pions in the described device, used in beam tests
of the CDF Plug Upgrade calorimeter \cite{Alb99}. Even with such simple devices,
pion rejection factors of the order of one hundred are readily achieved. 
Longitudinal segmentation of the calorimeter is thus most definitely {\bf not} not an essential requirement for this purpose.

Other reasons often used for longitudinal segmentation include the possibility to optimize the energy resolution of the em section, while limiting at the same time the cost of the hadronic section. However, in future experiments at the next generation high-energy lepton-lepton colliders, excellent energy resolution is needed for {\sl all} particles, not just electrons. Since sampling fluctuations are a major limiting factor both for electrons and hadrons in well designed dual-readout calorimeters, it stands to reason to use the same high sampling fraction and frequency throughout the calorimeter. This uniform structure is also a crucial factor for eliminating the intercalibration problems, illustrated in Sections 2.1.1 and 2.2.3, that plague {\sl all} longitudinally segmented non-compensating calorimeter systems \cite{Olga,Wig06b}.

Elimination of longitudinal segmentation also offers the possibility to make a finer lateral segmentation with the same number of electronic readout channels. This has many potential benefits. A fine lateral segmentation is crucial for recognizing closely spaced particles as separate entities. Because of the extremely collimated nature of em showers\footnote{Detailed measurements of the lateral profile of em showers in the RD52 calorimeter revealed a dominant central core with a diameter of only 3 mm \cite{RD52em}.}, it is also a crucial tool for recognizing electrons in the vicinity
of other showering particles. Moreover, a fine lateral segmentation is important for the identification of electrons in general. Unlike the vast majority of other calorimeter structures used in practice, the RD52 fiber calorimeter offers almost limitless possibilities for lateral segmentation. If so desired, one could read out every individual fiber separately. Modern silicon PM technology certainly makes that a realistic possibility.

\subsection{Other misconceptions}

Perhaps the most widespread misconception about calorimetry is the assumption that all calorimeter problems can be solved offline.
We are unaware of any convincing evidence in support of this assumption.


\section{Misconceptions deriving from beam tests of prototypes}

Before embarking on the construction of a calorimeter system, the (expected) performance is typically studied by exposing prototype modules to beams of different particles with different energies produced by an accelerator. Many of the misconceptions discussed in the previous section are the result of mistakes made in that process.

\subsection{The meaning of energy resolution}

One of the most important tasks of a calorimeter system is to measure the energy of particles or particle jets that are absorbed in it. The energy resolution is a measure of the calorimeter quality in this respect. The energy resolution is typically determined from the measured signal distribution for a beam of mono-energetic particles that enter the calorimeter in (approximately) the same impact point, usually the center of a module. However, one has to be careful interpreting the results of such measurements. We use experimental data obtained with the dual-readout lead/fiber calorimeter to illustrate this \cite{Lee17}.
\vskip 1mm
\begin{figure}[htbp]
\epsfysize=5cm
\centerline{\epsffile{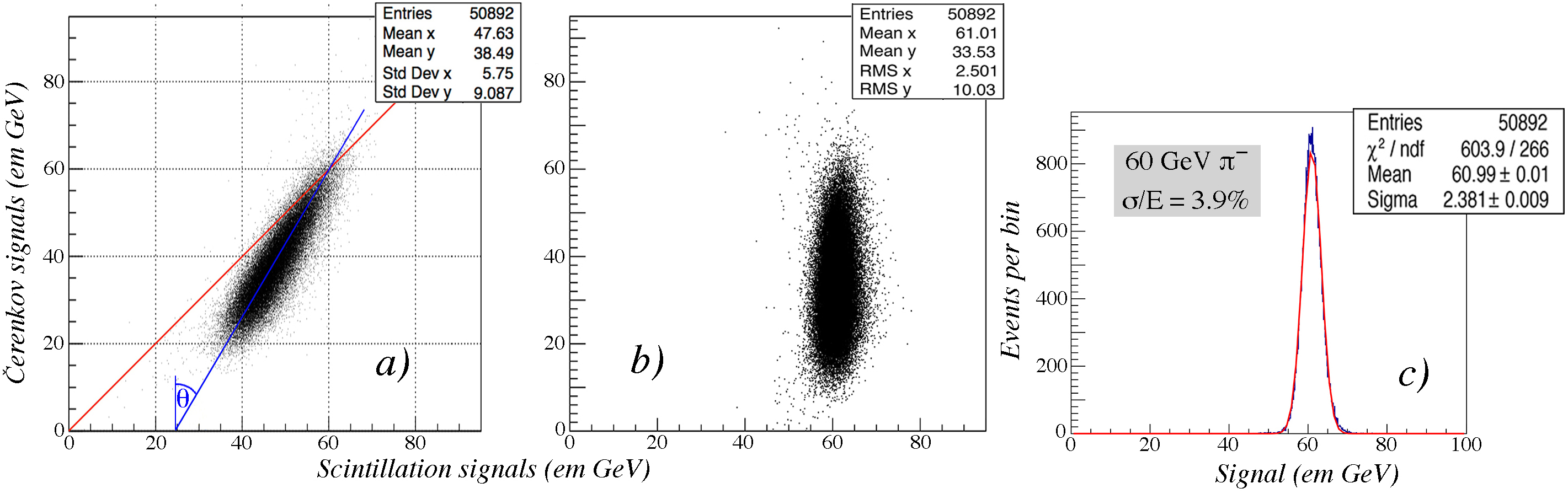}}
\caption{\footnotesize Signal distributions of the RD52 Dual-Readout lead/fiber calorimeter for 60 GeV pions \cite{Lee17}. Scatter plot of the two types of signals as recorded for these particles ($a$) and rotated around the point where the two lines from diagram $a$ intersect ($b$). Projection of the latter scatter plot on the $x$-axis ($c$).}
\label{DRrotation}
\end{figure}

Figure \ref{DRrotation}a shows a scatter plot of the \v{C}erenkov signals \vs the scintillation signals measured with this detector for 60 GeV pions. The calorimeter was laterally too small to fully contain the showers, which affected mainly the scintillation signals. In order to deal with this side leakage, the calorimeter was surrounded with 10 cm thick slabs of plastic scintillator material, and the signals from these counters were added to those from the scintillating fibers, using the fact that the measured shower profile indicated that the side leakage at this energy was, on average, 6.4\%. The energy scale for both the \v{C}erenkov and the scintillation signals is given in units of GeV, derived from the calibration of these signals with electron showers.

This scatter plot shows the data points located on a locus, clustered around a straight line that intersects the $C/S = 1$ line at the beam energy
of 60 GeV. This is to be expected \cite{PDG16}. In first approximation, the \v{C}erenkov fibers only produced signals generated by the electromagnetic components of the hadron showers, predominantly $\pi^0$s. The larger the em shower fraction, the larger the
$C/S$ signal ratio. Events in which (almost) the entire hadronic energy was deposited in the form of em shower components thus produced signals that were very similar to those from 60 GeV electrons and are, therefore, represented by data points located near (60,60) in this scatter plot.

We can now rotate the scatter plot over the angle $\theta$  around this intersection point,
%
%
the result is shown in Figure \ref{DRrotation}b. The projection of this rotated scatter plot on the $x$-axis is shown in Figure \ref{DRrotation}c.
This signal distribution is well described by a Gaussian function with a central value of 61.0 GeV and a relative width, $\sigma/E$, of $3.9\%$. This corresponds to $30\%/\sqrt{E}$. The narrowness of this distribution reflects the clustering of the data points around the axis of the locus in Figure \ref{DRrotation}a.\index{Energy resolution!measured with test beams}

The same procedure was applied for hadrons of other energies, covering a range from 20 - 125 GeV, and yielded similarly excellent results in terms of the response function. Interestingly, significant differences between pions and protons disappeared in this process. For example, at 80 GeV, the raw data showed that the average 
\v{C}erenkov signal was about 10\% larger for the pions than for the protons, confirming the effect shown in Figure \ref{pdif}.
However, using the intersection of the axis of the locus and the $C/S = 1$ point as the center of rotation, and the same rotation angle ($30^\circ$) as for 60 GeV, the resulting signal distributions had about the same average value: 80.7 GeV for the pions and 
80.4 GeV for the protons. The widths of both distributions were also about the same: 2.60 GeV for pions, 2.69 GeV for protons ($\sim 30\%/\sqrt{E}$).
Regardless of the differences between the production of $\pi^0$s (and thus of \v{C}erenkov light) in these two types of showers,
the signal distributions obtained after the dual-readout procedure applied here, were thus practically indistinguishable.

Yet, while we have managed to obtain very narrow signal distributions for the beam particles using only the calorimeter information, we don't think it is correct to interpret the relative width of these distributions as a measure for the precision with which the energy of an arbitrary particle absorbed in this calorimeter may be determined. The determination of the coordinates of the rotation point, and thus the energy scale of the signals, relied on the availability of {\sl an ensemble of events} obtained for particles of the same energy. In practice, however, one is only dealing with {\sl one} event and the described procedure can thus not be used in that case.
\vskip 1mm
\begin{figure}[b!]
\epsfysize=5.5cm
\centerline{\epsffile{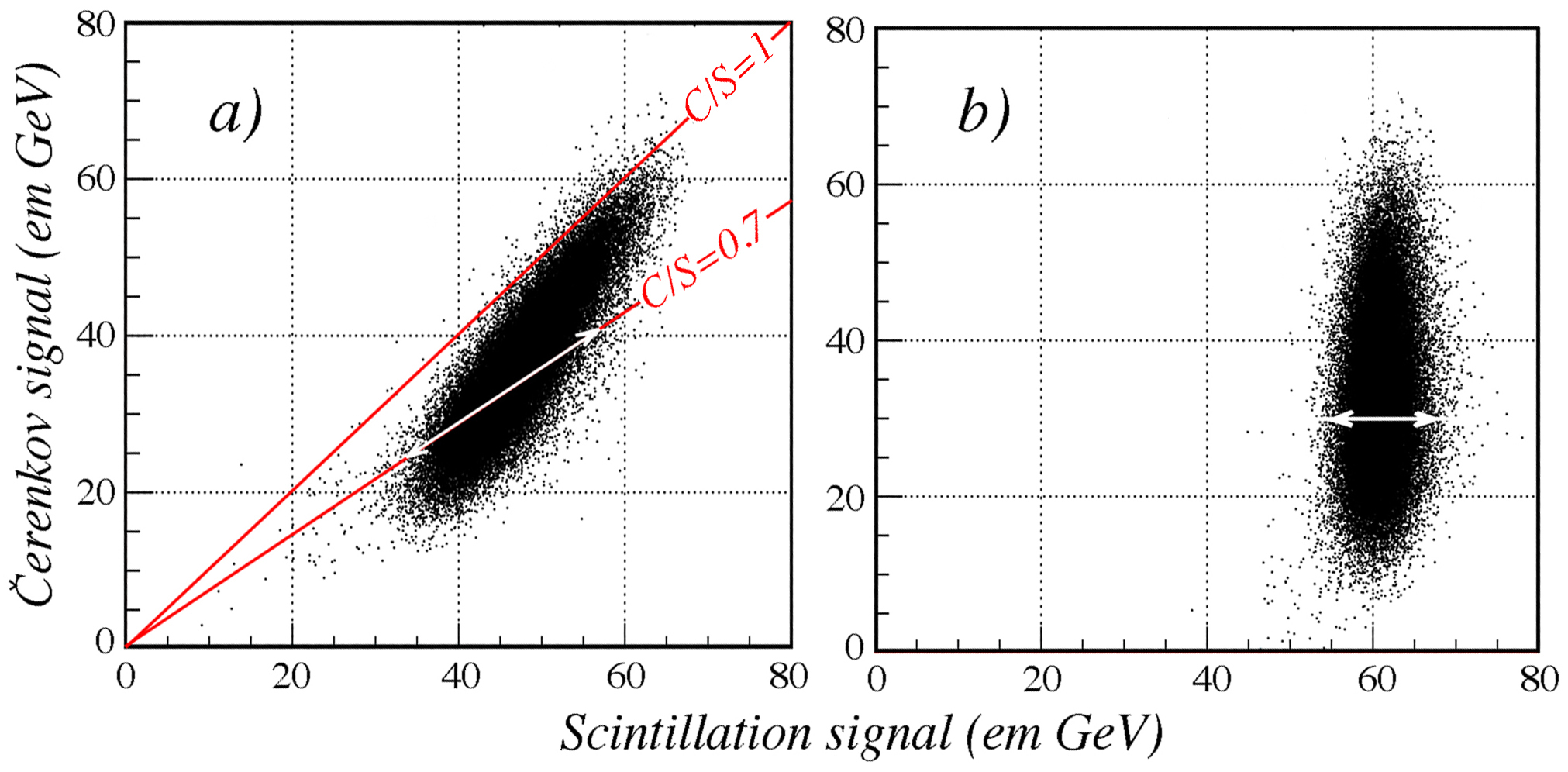}}
\caption{\footnotesize{Scatter plots of the \v{C}erenkov \vs the scintillation signals from showers induced by mono-energetic hadrons ($a$). 
The arrow indicates the precision with which the em shower fraction, and thus the energy, of an individual particle can be determined on the basis of the measured ratio of the \v{C}erenkov and scintillation signals, 0.7 in this example. The rotation procedure for an ensemble of 
mono-energetic pions leads to the scatter plot shown in diagram $b$.
The precision of the measurement of the width of that distribution is indicated by a white arrow as well \cite{Lee17}.}}
\label{dr2method}
\end{figure}

The DREAM Collaboration has developed a procedure to determine the energy of an unknown particle showering in the dual-readout calorimeter that is {\sl not} affected by this problem \cite{Akc05}. In this procedure, the em shower fraction ($f_{\rm em}$) of the hadronic shower is derived from the ratio of the \v{C}erenkov and scintillation signals. Using the known $e/h$ values of the two calorimeter structures, the measured signals can then be converted to the em energy scale ($f_{\rm em} = 1$). 
The energy resolutions obtained with this method are worse than the ones given above, although it should be mentioned that they are 
dominated by incomplete shower containment and the associated leakage fluctuations, and are likely to improve considerably for detectors that are sufficiently large. However, the same is probably true for the measurements of which the results are shown in Figure \ref{DRrotation}.
\vskip 2mm

The message we want to convey in this subsection is that one should not confuse the precision of the energy determination of a given event based on calorimeter signals alone with the width of a signal distribution obtained in a testbeam, since the latter is
typically based on additional information that is not available in practice. In the example described above, this additional information derived from the fact that a large number of events generated by particles of the same energy were available. In other cases, additional information may be derived from knowledge of the particle energy. This is especially true for calorimeters whose energy scale depends on ``offline compensation,'' or other techniques intended to minimize the total width of the signal distribution from a detector system consisting of several longitudinal segments. Such techniques depend on calibration constants whose values depend on the energy, on the type of showering particle, and sometimes also on the ratios of the signals from the different calorimeter sections. Moreover, the calibration constants also depend on the hadron type, and this leads to systematic
errors when protons (or kaons) are mistaken for pions. These errors were measured to be of the order of 4\% in the ATLAS Tilecal \cite{Adr09}. As shown in Figure \ref{pdif}, the response difference between pions and protons in the CMS Forward Calorimeter was measured to be $\sim 10\%$.

\subsection{Biased event samples}

One of the most common mistakes that are made when analyzing the performance of a calorimeter derives from the selections
that are made to define the experimental data sample. This selection process may easily lead to biases, which distort the performance characteristics
one would like to measure. In some extreme cases, this may lead to very wrong conclusions, such as the claim that uranium/liquid-argon calorimeters are 
compensating \cite{RWbook}. We illustrate this issue with data from the same calorimeter discussed in the previous subsection \cite{Lee17}. 
\vskip 1mm
\begin{figure}[htbp]
\epsfysize=9cm
\centerline{\epsffile{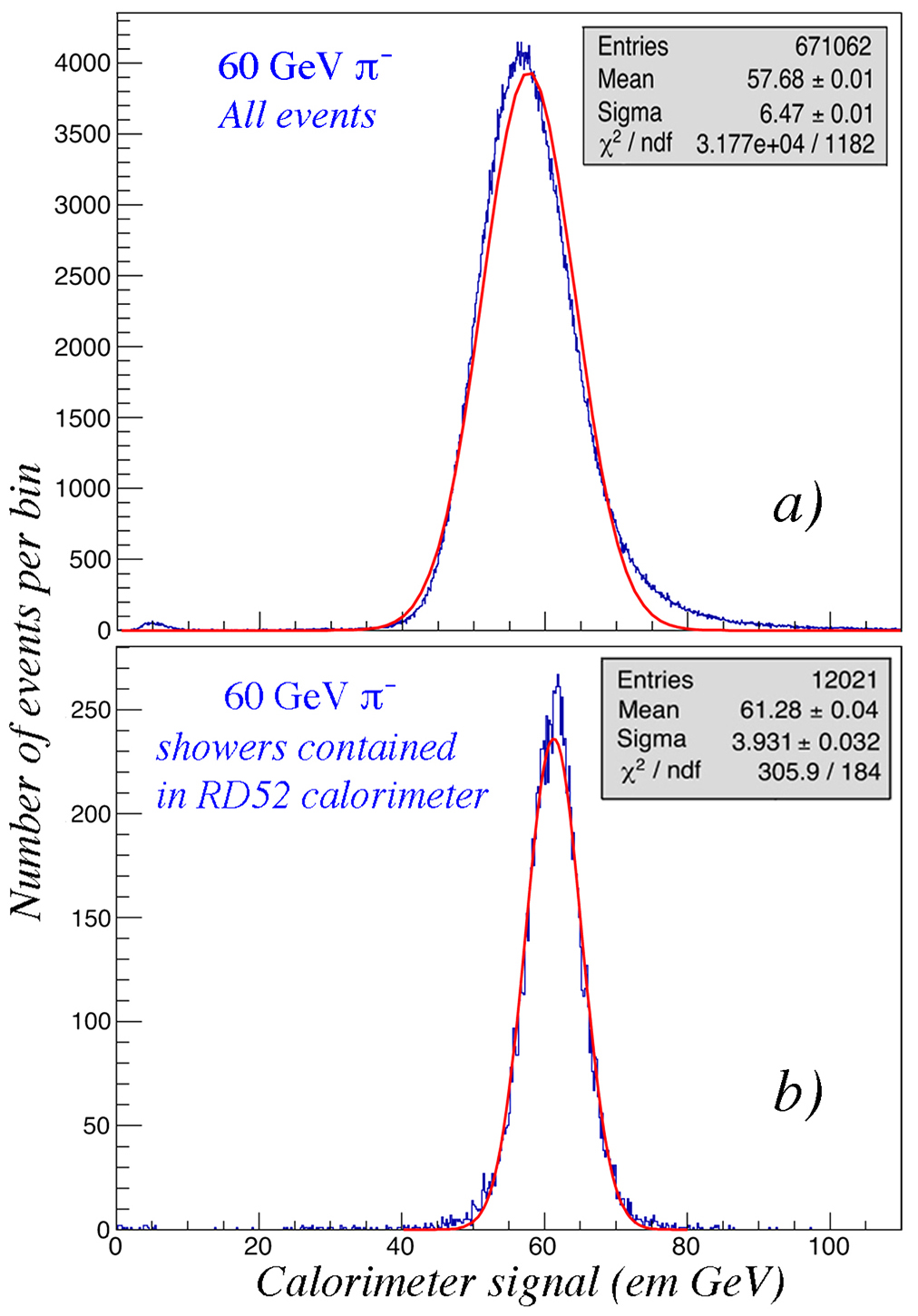}}
\caption{\footnotesize The signal distribution for 60 GeV $\pi^-$ in the RD52 dual-readout lead/fiber calorimeter, for all events ($a$) and for events where the showers were contained in this detector ($b$). See text for details.}
\label{pi60cont}
\end{figure}

Figure \ref{pi60cont} shows the signal distribution for 60 GeV $\pi^-$ in this calorimeter. The top diagram was obtained for {\sl all} events,
after applying the standard dual-readout procedure based on the measured scintillation and \v{C}erenkov signals. The bottom diagram shows the signal distribution for events in which the showers were fully contained in this calorimeter. Even though the leakage counters, which were used to select these events provided only partial coverage, the energy resolution was found to be more than 50\% better for this event sample. These results illustrate the fact that the hadronic energy resolution of this calorimeter was dominated by (lateral) shower leakage. Light attenuation in the (scintillating) fibers also contributed, as evidenced by the slightly asymmetric response function.

One could use this type of event selection to obtain spectacular performance in any calorimeter. For example, one could use a BGO crystal,
which serves as an em calorimeter in several experiments, and expose it to a test beam of pions. By requiring that the pions showers are fully contained in this crystal, one basically selects only charge exchange events, in which the pion converts its entire energy to a $\pi^0$ in the first interaction. The resulting signal distribution will look very much like that for electrons of the same energy as the pions. Of course, the selected event sample represents only a very small, non-representative, fraction of the total.

These examples illustrate the importance of {\sl unbiased} event samples in determining the performance of a calorimeter. As a general rule, the calorimeter data themselves should {\sl not} be used to apply cuts and thus select the event sample to be analyzed. Any such cuts should be based on external detectors.
Almost every analysis of test beam data we know of suffers from bias problems, the question is only {\sl to what extent} the obtained results are affected by this. Obviously, this may also make a comparison between results obtained with different calorimeters problematic, and less meaningful.

\section{Outlook}
\vskip 2mm

Calorimetry has come a long way in the past seventy years. Much of what has been learned about the inner workings of these somewhat mysterious detectors has been the result of dedicated generic research \& development projects, although the important contributions of work carried out in the ZEUS prototype phase definitely deserve an honorary mention in this context.

Unfortunately, times have changed. There are no longer significant resources available for this type of R\&D. New experiments are designed based on somebody's concept of what the detectors should look like, and prototype work primarily concentrates on technical aspects of that concept.
This approach is followed, for example, in projects carried out in the CALICE framework, which is geared towards application of PFA at future electron-positron colliders \cite{Sef15}.
Based on our observations, calorimetry research in this new era tends to be characterized by misconceptions (such as the ones discussed in the previous sections) and a general lack of interest in fundamental issues, combined with a strong belief that all eventual problems can be solved with technology. In this new paradigm, the use of tungsten, silicon, RPCs must of course lead to better results, because this is a more modern approach than lead + plastic.
\vskip 1mm
\begin{figure}[htbp]
\epsfysize=7cm
\centerline{\epsffile{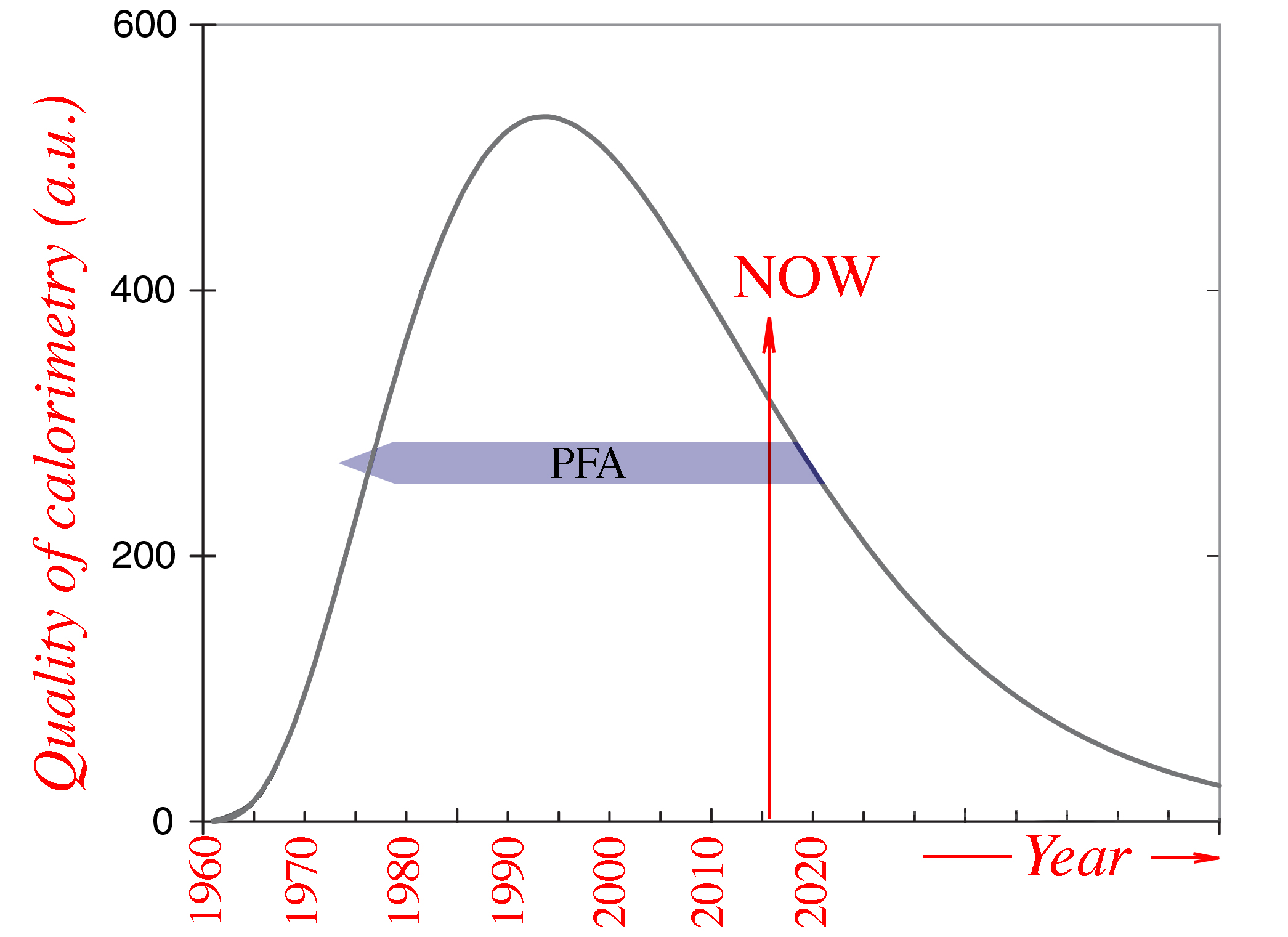}}
\caption{\footnotesize The quality of calorimetry for particle physics experiments as a function of time. See text for details.} 
\label{future}
\end{figure}

Because of these developments, we think the future of calorimetry is bleak. Figure \ref{future} depicts the quality of the calorimeters used in particle physics experiments as a function of time. This curve remarkably resembles a longitudinal shower profile. We have clearly passed the peak, and we are afraid that large-scale applications of PFA, such as envisaged in several existing and planned experiments, will set us back half a century, to the days of the magnetic spectrometers. PFA systems will actually be $4\pi$ versions of the magnetic spectrometers from those days, albeit on a scale that is an order of magnitude more compact (\eg 4 m instead of 40 m), with all the problems associated with that, since showers develop irrespective of where the absorbing calorimeters are located.
\newpage



\section*{References}
\begin{thebibliography}{10}

\bibitem{Liv95} Livan M, Vercesi V and Wigmans R (1995), {\em Scintillating-fibre 
Calorimetry}, CERN Yellow Report, CERN 95-02, Gen\`eve, Switzerland.
\bibitem{Olga} Albrow M \etal ~2002, \NIM {\bf A487}, 381.
\bibitem{Bern87} Bernardi E \etal ~1987, \NIM {\bf A262}, 229.
\bibitem{Aco91c} Acosta D \etal~1991, \NIM {\bf A308}, 481.
\bibitem{Suz99} Suzuki T \etal ~1999. \NIM {\bf A432}, 48.
\bibitem{Akc05} Akchurin N \etal ~2005, \NIM  {\bf A537}, 537.
\bibitem{Atl06} Aharrouche M \etal ~2006, \NIM {\bf A568}, 601.
\bibitem{Cer02} Cervelli F \etal ~2002, \NIM {\bf A490}, 132.   
\bibitem{Wig06b} Wigmans, R 2006, {\em Proc. of the workshop on hadronic shower simulation}, Fermilab (2006). AIP Conf. Proc. {\bf 896}, eds. M. Albrow and R. Raja (2007), 123. 
\bibitem{Sef15} Sefkow F \etal~2016, Rev. Mod. Phys. {\bf 88}, 1.
\bibitem{Pet12} Petyt D.A. 2012, \NIM {\bf A695}, 293.
\bibitem{Cih89} Cihangir S \etal~1989, \ieee {\bf NS-36}, 347.
\bibitem{Adl09} Adloff C \etal~2009, \NIM {\bf A608}, 372.
\bibitem{CMS07} Akchurin N \etal ~(CMS Collaboration)~2007, {\em The response of CMS 
combined calorimeters to single hadrons, electrons and muons}, CERN-CMS-NOTE-2007-012.
\bibitem{Ake87} \AA kesson T \etal ~1987, \NIM {\bf A262}, 243. 
\bibitem{Mehmet} Wigmans R and Zeyrek M 2002, \NIM {\bf A485}, 385. 
\bibitem{OUP2} Wigmans R 2017, {\em Calorimetry - Energy Measurement in Particle Physics}, 2nd ed., Oxford University Press, Chapter 6.
\bibitem{Tho09} Thomson M.A. 2009, \NIM {\bf A611}, 25.
\bibitem{Beh90} Behrens U \etal ~1990, \NIM {\bf A289}, 115.
\bibitem{Aba10} Abat E \etal~2010, \NIM {\bf A621}, 134.
\bibitem{You89} Young G.R. \etal~1989, \NIM {\bf A279}, 503.
\bibitem{Seg92} S\'eguinot J. \etal ~1992, \NIM {\bf A323}, 583.
\bibitem{Aad12} Aad G \etal~(ATLAS Collaboration) 2012,  \pl {\bf B716}, 1.
\bibitem{Chat12} Chatrchyan S \etal~(CMS Collaboration) 2012, \pl {\bf B716}, 30. 
\bibitem{deg07} Groom D.E. ~2007, \NIM {\bf 572}, 633.
\bibitem{Aco92b} Acosta D \etal ~1992, \NIM {\bf A316}, 184.
\bibitem{Akc98} Akchurin N \etal ~1998, \NIM {\bf A408}, 380.
\bibitem{pid} Akchurin N \etal ~2014, \NIM  {\bf A735}, 120.
\bibitem{Alb99} Albrow M \etal ~1999, \NIM {\bf A431}, 104.
\bibitem{RD52em} Akchurin N \etal ~2014, \NIM  {\bf A735}, 130.
\bibitem{Lee17} Lee S \etal~2017, {\sl Hadron detection with a dual-readout fiber calorimeter}, arXiv:1703.09120 [physics.ins-det]
\bibitem{PDG16} Patrignani C \etal ~2016 (Particle Data Group),  {\em Chin. Phys.} {\bf C40}, 100001, Section 34.9.2.
\bibitem{Adr09} Adragna P \etal ~2009, \NIM {\bf A606}, 362.
\bibitem{RWbook} Wigmans R 2017, {\em Calorimetry - Energy Measurement in Particle Physics}, 2nd ed., Oxford University Press, p. 640{\sl ff}.




\end{thebibliography}
\end{document}